\documentclass[twoside,11pt]{PhDthesisSU} %

\pdfoutput=1

\usepackage[T1]{fontenc} %
\usepackage[swedish,american]{babel}

\usepackage[backend=bibtex,style=authoryear,backref=false]{biblatex} %
\makeatletter

\newrobustcmd*{\parentexttrack}[1]{%
  \begingroup
  \blx@blxinit
  \blx@setsfcodes
  \blx@bibopenparen#1\blx@bibcloseparen
  \endgroup}

\AtEveryCite{%
  }

\makeatother
\usepackage{tocloft}


\let\cite\parencite
\let\citeN\textcite
\newcommand{\figuremacro}[3]{
	\begin{figure}[htbp]
		\centering
		\includegraphics[width=1\textwidth]{#1}
		\caption[#2]{\textbf{#2} - #3}
		\label{figure:#1}
	\end{figure}
}

\usepackage[autostyle]{csquotes}
\usepackage[normalem]{ulem} %
\usepackage{comment}
\usepackage{paralist} %
\usepackage[shadow,textsize=small]{todonotes}

\usepackage{pdfpages}
\usepackage{xspace}

\usepackage{longtable}

\newcommand{\ie}{\textit{i.e.,}\xspace}
\newcommand{\eg}{\textit{e.g.,}\xspace}
\newcommand{\sic}{\textit{sic}}
\newcommand{\viceversa}{\textit{vice versa}\xspace}

\usepackage{pifont}
\newcommand{\PLUS}{\texttt{+}\xspace}

\newcommand{\Y}{\ding{51}\xspace} %
\newcommand{\N}{\ding{55}\xspace} %
\newcommand{\R}{-\xspace}
\newcommand{\EQ}{$\vDash$\xspace}
\newcommand{\AND}{$\land$}

\newcommand{\IoT}{IoT\xspace}
\newcommand{\MMOW}{MMOW\xspace}
\newcommand{\FiiN}{FiiN\xspace}
\title{Distributed Technology-Sustained Pervasive Applications (v2.6)}
\subtitle{}
\author{\href{mailto:kim.nevelsteen@dsv.su.se}{Kim J.L. Nevelsteen}}

\begin{document}

\hypersetup{linkcolor=black}
\hypersetup{urlcolor=black}

\newcommand{\CNH}{CN:H\xspace}

\newcommand{\textTheGDD}{PAPER I}
\newcommand{\labelTheGDD}{thegdd}
	\newcommand{\TheGDD}{\textTheGDD\xspace}
	\newcommand{\citeTheGDD}{[\textTheGDD]\xspace}
\newcommand{\textAcrobats}{PAPER II}
\newcommand{\labelAcrobats}{acrobats}
	\newcommand{\Acrobats}{\textAcrobats\xspace}
	\newcommand{\citeAcrobats}{[\textAcrobats]\xspace}
\newcommand{\textTimeSpace}{PAPER III}
\newcommand{\labelTimeSpace}{timespace}
	\newcommand{\TimeSpace}{\textTimeSpace\xspace}
	\newcommand{\citeTimeSpace}{[\textTimeSpace]\xspace}
\newcommand{\textPervasiveMOO}{BOOK I}
\newcommand{\labelPervasiveMOO}{pervasivemoo}
	\newcommand{\PervasiveMOO}{\textPervasiveMOO\xspace}
	\newcommand{\citePervasiveMOO}{[\textPervasiveMOO]\xspace}
\newcommand{\textWorldDEF}{PAPER VI}
\newcommand{\labelWorldDEF}{worlddef}
	\newcommand{\WorldDEF}{\textWorldDEF\xspace}
	\newcommand{\citeWorldDEF}{[\textWorldDEF]\xspace}
\newcommand{\textDelegation}{PAPER VII}
\newcommand{\labelDelegation}{delegation}
	\newcommand{\Delegation}{\textDelegation\xspace}
	\newcommand{\citeDelegation}{[\textDelegation]\xspace}

\selectlanguage{american}

\frontmatterSU

\halftitlepage

\maketitle  %

\newpage
\thispagestyle{empty}

\newcommand{\DOMAIN}{}
\newcommand{\PROBLEM}{}
\newcommand{\RQ}{}
\newcommand{\RELATEDWORK}{}
\newcommand{\APPROACH}{}
\newcommand{\VERIFICATION}{}
\newcommand{\CONCLUSION}{}
\newcommand{\IMPLICATIONS}{}

\clearpage

\phantom{.}

\vspace{\stretch{1}}

{\fontfamily{verdana}\selectfont
{\scriptsize
\noindent
Kim J.L. Nevelsteen, Stockholm University, 2015. %
 
\vspace{5mm}
\noindent
ISBN 978-91-7649-277-2\\ %
ISSN 1101-8526\\ 
DSV Report Series No. 15-016\\

\noindent
Printer: Holmbergs, Malm\"{o} 2015\\
Distributor: Department of Computer and Systems Sciences

\noindent
Cover image: rendition of \cite[Figure 2]{lankoski2004-songsnorth}, see pp.~\pageref{figure:p413-lankoski-Figure-2}
}
}

\cleardoublepage


\begin{abstracts} 
{\small


\DOMAIN 
Technology-sustained pervasive games, contrary to technology-supported pervasive games, can be understood as \textit{computer games interfacing with the physical world}. 
Pervasive games are known to make use of `non-standard input devices' 
and with the rise of the \mbox{Internet of Things} (\IoT), 
pervasive applications can be expected to move beyond games.
This dissertation is requirements- and development-focused \mbox{Design Science} research for distributed technology-sustained pervasive applications, incorporating knowledge from the domains of Distributed Computing, Mixed Reality, Context-Aware Computing, Geographical Information Systems and \IoT.
\PROBLEM
Computer video games have existed for decades, with a reusable game engine to drive them.
\RQ If pervasive games can be understood as computer games interfacing with the physical world, can computer game engines be used to stage pervasive games? Considering the use of non-standard input devices in pervasive games and the rise of \IoT, how will this affect the architectures supporting the broader set of pervasive applications?
\RELATEDWORK
The use of a game engine can be found in some existing pervasive game projects, but general research into how the domain of pervasive games overlaps with that of video games is lacking. When an engine is used, a discussion of, what type of engine is most suitable and what properties are being fulfilled by the engine, is often not part of the discourse.
\APPROACH
This dissertation uses multiple iterations of the method framework for Design Science for the design and development of three software system architectures. In the face of \IoT, the problem of extending pervasive games into a fourth software architecture, accommodating a broader set of pervasive applications, is explicated.
\VERIFICATION
The requirements, for technology-sustained pervasive games, are verified through the design, development and demonstration of the three software system architectures. The scaling up of the architecture to support distributed pervasive applications, is based on research in the domain of Virtual Worlds and \IoT.
\CONCLUSION
The results of this dissertation are: the aligning of the Pervasive Games research domain with that of Virtual Worlds, the mapping of virtual time and space to physical counterparts, the scaling up of pervasive games to distributed systems, and the explication of the problem of incorporating \IoT into pervasive applications.
\IMPLICATIONS
The implication of this dissertation is to ensure that pervasive games are not left reinventing existing technologies.

}
\end{abstracts}


 %

%
%



\selectlanguage{swedish}


\chapter*{Sammanfattning}
{\small


Teknikf\"ormedlade verklighetsspel (technology-sustained pervasive games), i motsats till teknikst\"odda verklighetsspel (technology-supported pervasive games), kan f\"orst\r{a}s som \textit{dataspelets gr\"anssnitt mot den fysiska v\"arlden}. Verklighetsspel games \"ar k\"anda f\"or att anv\"anda sig av `icke-standardiserade inmatningsenheter' och med \"okningen av \mbox{Sakernas Internet (Internet of Things)} (\IoT), kan verklighetsapplikationer (pervasive applications) f\"orv\"antas g\r{a} l\"angre \"an verklighetsspel. Denna avhandling omfattar krav- och utvecklingfokuserad (Design Science) forskning f\"or distribuerad teknik omfattande verklighetsspel, som inneh\r{a}ller kunskap fr\r{a}n omr\r{a}dena distribuerad databehandling (Distributed Computing), blandad realitet (Mixed Reality), kontextmedveten databehandling, geografiska informationssystem och \IoT. 
Dataspel har funnits i decennier, ofta med en \r{a}teranv\"andbar spelmotor f\"or att driva dem. Om verklighetsspel kan f\"orst\r{a}s som dataspel med gr\"anssnitt mot den fysiska v\"arlden, kan d\r{a} dataspelsmotorer anv\"andas f\"or att iscens\"atta verklighetsspel? Med tanke p\r{a} anv\"andningen av icke-standardiserade inmatningsenheter i verklighetsspel och den tilltagande m\"angde \IoT till\"ampningar, hur kommer detta att p\r{a}verka arkitekturen som st\"oder verklighetsspel? Anv\"andningen av en konventionell spelmotor kan \r{a}terfinnas i vissa befintliga verklighetsspelsprojekt, men mer generell forskning om hur verklighetsspel \"overlappar med konventionella dataspel saknas. N\"ar en konventionell dataspelsmotor anv\"ands, \"ar en diskussion om vilken typ av motor som \"ar mest l\"amplig och vilka egenskaper uppfylls av motorn ofta inte en del av diskursen. Denna avhandling anv\"ander flera iterationer av metodramverket f\"or design vetenskap (method framework for Design Science) f\"or konstruktion och utveckling av tre mjukvarusystemarkitekturer. Med tanke p\r{a} \IoT utarbetas problemet att utvidga verklighetsspel till en fj\"arde mjukvaruarkitektur som kan tillm\"otesg\r{a} en bredare upps\"attning av verklighetsapplikationer. 
Kraven f\"or teknikf\"ormedlade verklighetsspel verifieras genom design, utveckling och demonstration av tre mjukvarusystemarkitekturer. Uppskalning av arkitekturen f\"or att st\"odja distribuerade verklighetsspel \"ar baserad p\r{a} forskning inom omr\r{a}det f\"or virtuella v\"arldar och \IoT. Resultaten fr\r{a}n avhandlingen \"ar: anpassning av forskningsomr\r{a}det verklighetsspel med forskningsomr\r{a}det virtuella v\"arldar, metod f\"or matchning av virtuell tid och utrymme till fysiska motsvarigheter, uppskalning av verklighetsspel till distribuerade system, och utarbetning av problemen med att inf\"orliva \IoT in verklighetsapplikationer. Inneb\"orden av denna avhandling \"ar att se till att implementeringen av verklighetsspel inte leder till att man \r{a}teruppfinner redan existerande teknik.


}
\selectlanguage{american}


\cleardoublepage



\begin{dedication}

{\fontfamily{calligra}\selectfont
{\LARGE
Dedicated to \dots
}}

\end{dedication}

%



\chapter*{Acknowledgements}

\hypersetup{linkcolor=blue}
\hypersetup{urlcolor=blue}

\chapter{List of Publications}
\label{ch:publist}

\vspace{-5pt} %

The following papers, referred to in the text by their Roman numerals, are included in this thesis. 

\vspace{0pt} %
\bigskip

\noindent\begin{longtable}{ll}

\TheGDD : &	
	\begin{minipage}[t]{0.8\columnwidth}%
	\textbf{GDD as a Communication Medium}\\
	\citeN{nevelsteen2012-gdd}
	\end{minipage}\tabularnewline\tabularnewline

\Acrobats : &	
	\begin{minipage}[t]{0.8\columnwidth}%
	\textbf{Athletes and Street Acrobats:\\ Designing for Play as a Community Value in Parkour}\\
	\citeN{waern2012-acrobats}
	\end{minipage}\tabularnewline\tabularnewline

\TimeSpace : &	
	\begin{minipage}[t]{0.8\columnwidth}%
	\textbf{Spatiotemporal Modeling of a Pervasive Game}\\
	\citeN{nevelsteenDRAFT-spatiotemporal}
	\end{minipage}\tabularnewline\tabularnewline

\begin{minipage}[t]{0.2\columnwidth}%
\PervasiveMOO : \\ 
(PAPER IV \\
\&~PAPER V)
\end{minipage} &
	\begin{minipage}[t]{0.8\columnwidth}%
	\textbf{A Survey of Characteristic Engine Features for\\ Technology-Sustained Pervasive Games}\\
	\citeN{nevelsteen2015-pervasivemoo}
	\end{minipage}\tabularnewline\tabularnewline

\WorldDEF : &	
	\begin{minipage}[t]{0.8\columnwidth}%
	\textbf{`Virtual World', Defined from a Technological Perspective, and Applied to Video Games, Mixed Reality and\\ the Metaverse}\\
	\citeN{nevelsteenDRAFT-virtualworlddef}
	\end{minipage}\tabularnewline\tabularnewline

\Delegation : &	
	\begin{minipage}[t]{0.8\columnwidth}%
	\textbf{Comparing Properties of Massively Multiplayer Online Worlds and Internet of Things}\\
	\citeN{nevelsteenDRAFT-delegation}
	\end{minipage}\tabularnewline\tabularnewline
	
\end{longtable}

\noindent
\rule{\linewidth}{0.5mm}

\newpage

\section*{Not Included in this Dissertation}

\noindent\begin{longtable}{ll}

PAPER : &	
	\begin{minipage}[t]{0.8\columnwidth}%
	\textbf{Applying GIS Concepts to a Pervasive Game: \\ Spatiotemporal Modeling and Analysis Using the\\ Triad Representational Framework}\\
	Kim J. L. Nevelsteen (2014).
	\emph{International Journal of\\ Computer Science Issues} (IJCSI), \textbf{11}(5). \\ %
	E-ISSN:~\href{http://ijcsi.org/articles/Applying-gis-concepts-to-a-pervasive-game-spatiotemporal-modeling-and-analysis-using-the-triad-representational-framework.php}{1694-0814} (NSD niv\aa\ 1, 2014) %
	\end{minipage}\tabularnewline\tabularnewline

\end{longtable}

\noindent
\rule{\linewidth}{0.5mm}

\vspace{2mm}

\noindent
Reprints were made with permission from the publishers.

\section*{Author's Contributions}
\label{ch:pubcontrib}

The contribution of each published work is summarized in the following list:

\subsection*{\textTheGDD}

\subsubsection*{\textit{GDD as a Communication Medium}}

\noindent
Requirements research and the original architectural design of the GDD was done by this author. \TheGDD is done in collaboration with Sergio Gayoso Fern\'{a}ndez; a master student under supervision at the time. 
With the help of this author, Sergio Gayoso adapted the architectural design, in three iterations, including the staged case studies and phenomenological interviews. The GDD proved to model a persistent communication technology, of which the requirements analysis, design and the notion of a `view' are a contribution. Collaborative user editing via the Internet was not mainstream at the time and the `view' concept is still not wide spread today.

\subsection*{\textAcrobats}

\subsubsection*{\textit{Athletes and Street Acrobats:\\ Designing for Play as a Community Value in Parkour}}

The design of a pervasive service to satisfy the project stakeholders and the Parkour community included: the design of an architecture to sustain the pervasive service and a meet-up map function. The design of the service itself was not done by this author, but by others in the research group. The design of the architecture was done by Annika Waern (our research group) and Joel Westerberg (Street Media 7), with added input from this author. Design of the map functionality was done solely by this author, in collaboration with the author's supervisor at the time, Annika Waern. Final redesign and trials of the service were done by Elena Balan, under the supervision of Annika Waern and with the help of this author.

Traveur, of \Acrobats, highlights that ``heterogeneity is the norm'' \ie that having both heterogeneous servers and clients is become far more common. Traveur served as exploratory research into pervasive and context-aware computing, of which the requirements analysis and design are a contribution; poor architectural design lead to broadening the search for technologies to included solutions from the video game industry. 

\subsection*{\textTimeSpace}

\subsubsection*{\textit{Applying GIS Concepts to a Pervasive Game: Spatiotemporal Modeling and Analysis Using the Triad Representational Framework}}

This author is the sole author of this work. 
In the domain of Pervasive Games, research exists tying the virtual to the physical, but an ``integrated model for dealing with space and time''~\citeTimeSpace is lacking. Such a model does exist in the domain of GIS, and since the domains of GIS and Pervasive Games overlap (based on Earth's geography) the Triad Representational Framework can be exapted to pervasive games. The contribution of \TimeSpace is the demonstration that Triad is indeed applicable to pervasive games.

\subsection*{\textPervasiveMOO}

\subsubsection*{\textit{A Survey of\\ Characteristic Engine Features for Technology-Sustained Pervasive Games}}

This author is the sole author of this work. Chapter 2 of \PervasiveMOO is an extensive systematic review into pervasive games. The resulting feature set is a substantial contribution describing characteristic features of a would-be pervasive games engine. These features can be considered a set of informal requirements from which a set of formal requirements can be drawn. Using the feature set, a virtual world engine was chosen as being in the same \mbox{`product line'}~\cite{bass2013-inpractice} as a would-be pervasive games engine, based on the shared trait of a persistence. In Chapter 3 of \mbox{\PervasiveMOO}, the component feature set and the choice of a virtual engine as pervasive engine are verified through the case study of the pervasive game called CN:H. \PervasiveMOO highlighted that a model, mapping time and space from the physical to the virtual, and \viceversa, was missing and that no usable definition for a `virtual world' existed.

\subsection*{\textWorldDEF}

\subsubsection*{\textit{`Virtual World', Defined from a Technological Perspective, and Applied to Video Games, Mixed Reality and
the Metaverse}}

This author is the sole author of this work. The obvious contribution of \WorldDEF is a definition for a `virtual world' from a technological perspective, rather than a cultural one, with all underlying properties of the definition defined in detail as well. Less explicit contributions are: the determining what constitutes a virtual world when dealing with a distributed and possibly disconnected architectures; how a virtual world relates to virtual and mixed reality; and, an ontology showing the relationship between complimentary terms and acronyms. To verify the definition, it is used to classify contemporary technologies \eg Destiny.

\subsection*{\textDelegation}

\subsubsection*{\textit{Comparing Properties of Massively Multiplayer Online Worlds and\\ Internet of Things}}

Research and drafting of property requirements relating Massive Multiplayer Online Worlds and the Internet of Things is done by this author. Conceptual work on decentralizing a game engine for use with Internet of Things is done by this author, in collaboration with supervisors, Theo Kanter and Rahim Rahmani. Each of the case studies is provided by a single author; MediaSense by Theo Kanter, Immersive Networking by Rahim Rahmani and Virtual Worlds as a ``behind the scenes'' resource by this author.

\Delegation surveys the domain of {\MMOW}s (large scale virtual worlds) for properties that are affected by scaling an architecture, and then evaluates how these are dealt with in the domain of IoT. The contribution of \Delegation is the problem explication of scaling architectures for {\MMOW}s and IoT; six properties related to scaling are discussed. That device and system heterogeneity is an issue in the domains of Virtual Worlds and \IoT, is particularly relative to this dissertation.

\hypersetup{linkcolor=black}
\hypersetup{urlcolor=black}

\cleardoublepage

\setcounter{secnumdepth}{3} %
\setcounter{tocdepth}{3}    %
\tableofcontents            %

\renewcommand{\nomname}{Abbreviations} %

\begin{footnotesize} %

\printnomenclature[2 cm] %
\label{nom} %

\end{footnotesize}

\cleardoublepage

\renewcommand\cftloftitlefont{\huge}
\renewcommand\cftlottitlefont{\huge}

\listoffigures	%

\listoftables  %

\hypersetup{linkcolor=blue}
\hypersetup{urlcolor=blue}

\cleardoublepage

\pagenumbering{gobble}
\part{Thesis}

\mainmatterSU

\graphicspath{{includes/1_img/}} %

\chapter{Introduction} %
\label{ch:introduction}

Pervasive games are becoming more common place \eg with Google, the multinational company, staging a pervasive game called Ingress~\cite{niantic2013-ingress}, and apps like Zombies, Run!~\cite{sixtostart2012-zombies} becoming more mainstream.
According to \citeN{benford2005-bridging}, ``pervasive games extend the gaming experience out into the real [physical] world''. 
A pervasive game according to the definition by \citeN{montola2005-pgdef} ``is a game that has one or more salient features that expand the contractual magic circle of play socially, spatially or temporally'' \ie ``expand the boundaries of play''~\cite{oppermann2009-lbxp}. In her definitive work, \citeN[original italics]{nieuwdorp2007-discourse} derives that pervasive games can be discussed from two perspectives, a technological one, ``that focuses on computing technology \textit{as a tool to enable the game to come into being}'' or a cultural one, ``that focuses on \textit{the game itself}''. 
There is a class of pervasive games which are `technology-sustained', relying on computer simulation to maintain game state and react to player activities; these games can be understood as ``computer games interfacing with the physical world''~\cite[p.164]{montola2009}. Technology-sustained pervasive games are contrary to `technology-supported' games, where not all game activities are supported by information technology~\cite{montola2009} \ie do not necessarily require a game engine.
Pervasive games are known to make use of `non-standard input devices'~\cite{nieuwdorp2007-discourse} and with the rise of the \mbox{Internet of Things (\IoT)}~\cite{gartner2014-iot}, pervasive applications can be expected to move beyond games.
This dissertation is requirements- and development-focused Design Science research~\cite[p.79]{johannesson2014-designscience} for distributed technology-sustained pervasive applications, incorporating knowledge from the domains of Distributed Computing, Mixed Reality, Context-Aware Computing, Geographical Information Systems (GIS) and \IoT.

\section{Problem Statement}

Computer video games have existed for decades, with reusable game engines to drive them; the major incentive for employing a reusable game engine being reduced development time and cost~\cite{lewis2002,bass2013-inpractice}. Currently, there are no reusable game engines available for pervasive games; without such engines, developers would be left continually reinventing existing technologies. If technology-sustained pervasive games can be understood as computer games interfacing with the physical world, can computer game engines be used to stage a pervasive game? And, can advances from the domain of Computer Video Games to be used to scale up pervasive games to distributed systems? To answer these questions, a requirements analysis for pervasive games must be performed, an architecture found that correlates to those requirements, and an analysis done to verify if existing architectural properties actually overlap with the requirements. 
According to \citeN{jonsson2007}, pervasive games need a sensory system to monitor the physical world; access to \IoT could potentially serve as such a sensory system. Considering the use of non-standard input devices in pervasive games and the rise the \IoT, how will this affect pervasive games architectures, allowing for a broader set of pervasive applications?

\section{Research Question}

%
%

%

Given the domain and the problem statement, the research question for this dissertation is then:

\blockquote{\textit{To allow advances from the domain of Computer~Video~Games to be used to scale up pervasive games to distributed systems, can engine technology from the domain of Computer Video Games be repurposed to stage technology-sustained pervasive games? Considering the use of non-standard input devices in pervasive games and the rise of Internet of Things, how will this affect the \mbox{architectures} supporting the broader set of pervasive applications?}
}

\noindent 
These research questions are answered by first doing a requirement analysis into pervasive games, and then expanding the requirements to pervasive applications. Advances in the other domains, mentioned above, are taken into account during the analysis and design. Three software system architectures are created in four iterations to verify the requirements and provide input for further research. 

\section{Research Approach}
\label{section:research_approach}

This dissertation incorporates knowledge from the domains of Distributed Computing, Mixed Reality, Context-Aware Computing, GIS and \IoT. Because the solution maturity in the other domains (\eg Computer Video Games and GIS) is high and the application domain maturity of Pervasive Games is low, this dissertation makes use of several exaptations; an `exaptation' being the extension of a known solution to a new problem~\cite[p.11]{johannesson2014-designscience}. 
Four iterations of the method framework for Design Science~\cite{johannesson2014-designscience} are used for the design and development of three software system architectures: the Game Design Document (GDD), a new game design and communication medium; Traveur, a pervasive game turned pervasive service due to liquid requirements; and, Codename: Heroes (\CNH), a ``long-term pervasive game''.
In the face of \IoT, the fourth iteration, explicates the problem, of extending pervasive games into a broader set of pervasive applications.

\begin{enumerate}[I:]

\item The first architecture was a design exercise on how to model a GDD medium, in the form of a `living document' (see Section~\ref{section:architecture-thegdd}). In hindsight, the project proved to be an exploratory case study; the architecture, which proved to be a persistent communication technology (see Section~\ref{section:gdd-vs-virtual-world}), and the `view' concept, influenced subsequent iterations. Results of this iteration can be found in \TheGDD.

\item In this iteration, an architecture for Traveur was designed, a pervasive service (originally a pervasive game) and a map function for that service. Publication of the architecture was created in draft, but never published; to compensate, it is discussed in depth below (see Section~\ref{section:architecture-traveur}). Redesign of the service itself and the map function was published in \Acrobats. This project exemplifies shifting design requirements, heterogeneity as the norm, and context-awareness.

\item The most extensive architecture herein, for \CNH, was with the aim of designing an architectural back-end for a technology-sustained pervasive game (see Section~\ref{section:architecture-pervasivemoo}). 
Chapter~2 of \PervasiveMOO contains a requirements analysis, through a literature review, for pervasive games; Chapter~3 contains a demonstration of the requirements and resulting architecture as case study; and Chapter~4 is a summary of the challenges and open issues for pervasive games, which influenced the final iteration IV.

\item A virtual world that supports a massive number of players, often referred to as a Massively Multiplayer Online World (\MMOW), are often enabled by distributed architectures. \Delegation explicates the problem of scaling architectures for a \MMOW and \IoT. 
Because the domain of Pervasive Games is aligned with the domain of Virtual worlds herein, results from \Delegation are combined with the open issues from \PervasiveMOO, so as to explicate the problem of extending pervasive games into a broader set of pervasive applications incorporating \IoT. This then leads into the future work of Chapter~\ref{ch:futurework}.

\end{enumerate}

\noindent
Having completed \CNH, in Iteration III, it was clear that a model mapping the virtual to the physical (and \viceversa) was lacking, as well as a definition for `virtual world', not withstanding that so called `virtual world engines' exist. In \mbox{\TimeSpace}, a solution for mapping the virtual and physical is borrowed from the domain of GIS and applied to pervasive games. And, a definition for a `virtual world' is obtained using grounded theory in \WorldDEF. Because, in \PervasiveMOO, research in pervasive games is aligned with that of virtual worlds (based on the persistence trait), concepts in \WorldDEF (\eg mixed reality and distributed computing) could be subsequently applied to pervasive games.

\section{Contributions}

In its entirety this dissertation provides \textit{requirements for technology-sustained pervasive games} and an \textit{explicated problem, from which requirements can be formed, for distributed pervasive applications}; requirements are verified by designs and feasibility implementations. In addition to the contributions found in each of the individual publications (see the Author's Contributions~p.~\pageref{ch:pubcontrib}), the dissertation contributions include the following points:

\begin{itemize}
\renewcommand{\labelitemi}{$\square$}

\item Containing the design and development of three software system architectures and clearly showing how each influenced the requirements for technology-sustained pervasive games; the requirements analysis and design of the GDD and Traveur influencing, including game master interfaces, heterogeneous servers, context-awareness and mixed reality.

\item Using \mbox{\WorldDEF}, it is shown that domain of Pervasive Games overlaps with that of Virtual Worlds, except for a discrepancy based on mixed reality; %
that the majority of properties defining a virtual world are characteristics that are preferable to a would-be pervasive games engine \ie a substantial improvement over choosing a virtual world engine based solely on the overlapping trait of persistence, as in ~\PervasiveMOO.

\item To handle the mixed reality nature of pervasive games, the Triad Representational Framework~\citeTimeSpace is combined with \cite{dix2005} and \cite{langran1992-timegis} to map time and space, from the virtual to the physical, and \viceversa; the combined result can be directly implemented in engine technology. 

\item By aligning pervasive games with that of virtual worlds, advances in architecture, from the domain of Virtual Worlds, are used to scale up pervasive games architectures to distributed systems. %

\item Using results from \Delegation pervasive games are extended to pervasive applications, incorporating \IoT; properties pertaining to scalability from the domain of {\MMOW}s and IoT are compared and the result used to explicate the problem of scaling architectures for pervasive applications.

\item The domain of Distributed Pervasive Applications is linked to the domains of Distributed Computing, Mixed Reality, Context-Aware Computing, GIS and \IoT, see Figure~\ref{figure:overlapping-domains}. \TheGDD, \mbox{\WorldDEF}, \mbox{\PervasiveMOO} and \Delegation all highlight challenges surrounding Distributed Computing. \mbox{\WorldDEF} ties Virtual Worlds to Mixed Reality, and \PervasiveMOO, with the help of \mbox{\WorldDEF}, shows Pervasive Games to overlap with both the domain of Virtual Worlds and Mixed Reality. Traveur, the pervasive service in \Acrobats, makes heavy use of Context-Aware Computing, and \PervasiveMOO links this to Pervasive Games. \TimeSpace ties the domain of GIS to Pervasive Games based on Earth's geography. And, \mbox{\Delegation} expands the concept of pervasive games to pervasive applications, tying all the above to the domain of \IoT. 

\bigskip
\bigskip
\begin{minipage}{\linewidth}
	\centering
	\includegraphics[width=\linewidth]{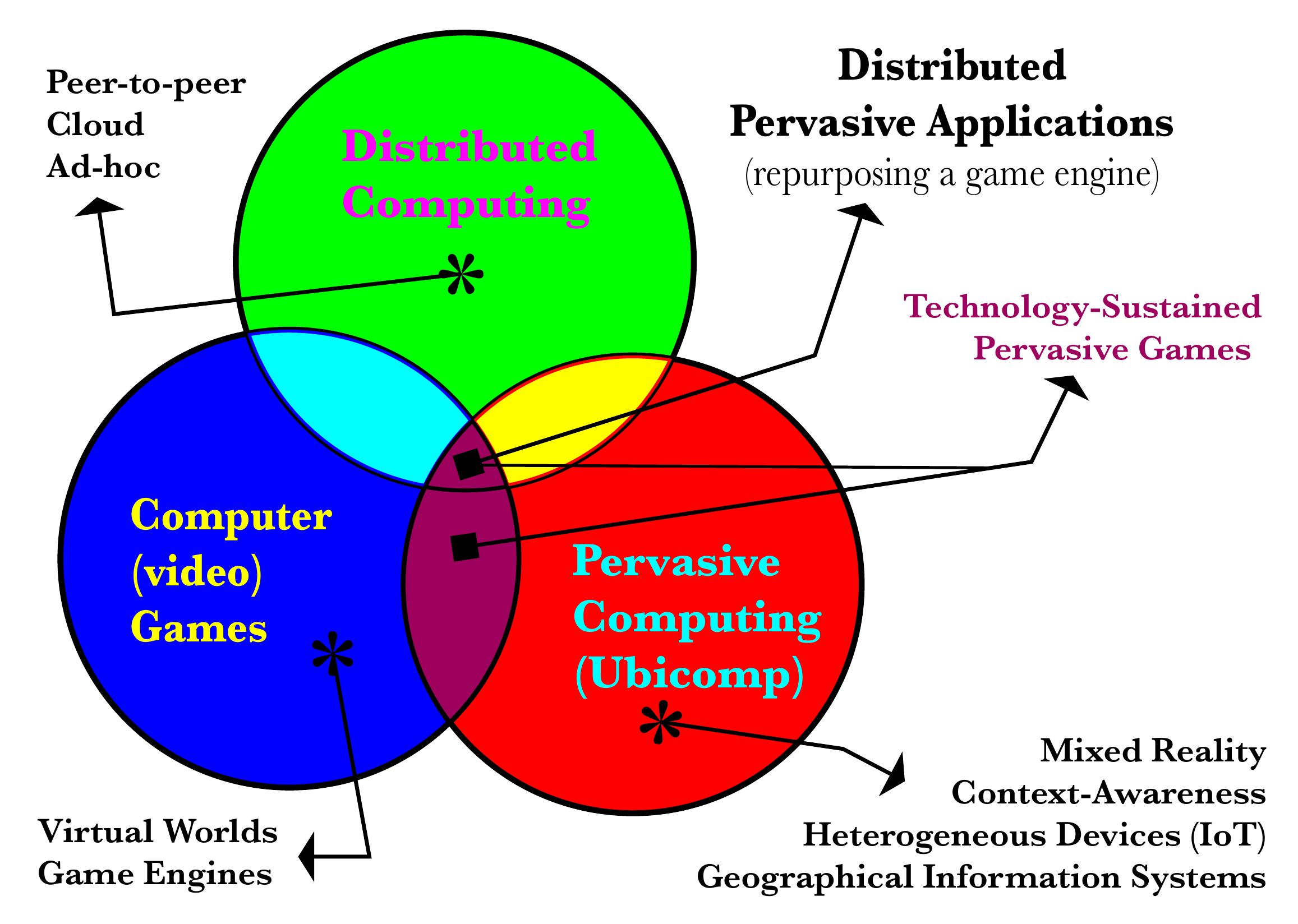}
	\captionof{figure}{Overlapping domains of Distributed Pervasive Applications %
}
	\label{figure:overlapping-domains}
	\end{minipage}
\end{itemize} %

\newpage
\section{Dissertation Structure}

This dissertation is structured such that: The Introduction, you have just read. Research Strategies and Methods are discussed in Chapter~\ref{ch:methodology}, with one underlying section for each iteration of the method framework. Related work is presented in Chapter~\ref{ch:relatedwork}. Chapter~\ref{ch:architectures} contains three software system architectures in three iterations, one section for each iteration. In Chapter~\ref{ch:combinedwork} the mixed reality of pervasive games is dealt with and architectures are scaled up to distributed systems; and a section (Section~\ref{section:delegation}) for the final Iteration IV, incorporating \IoT (see Figure~\ref{figure:dissertation-roadmap}).

\figuremacro{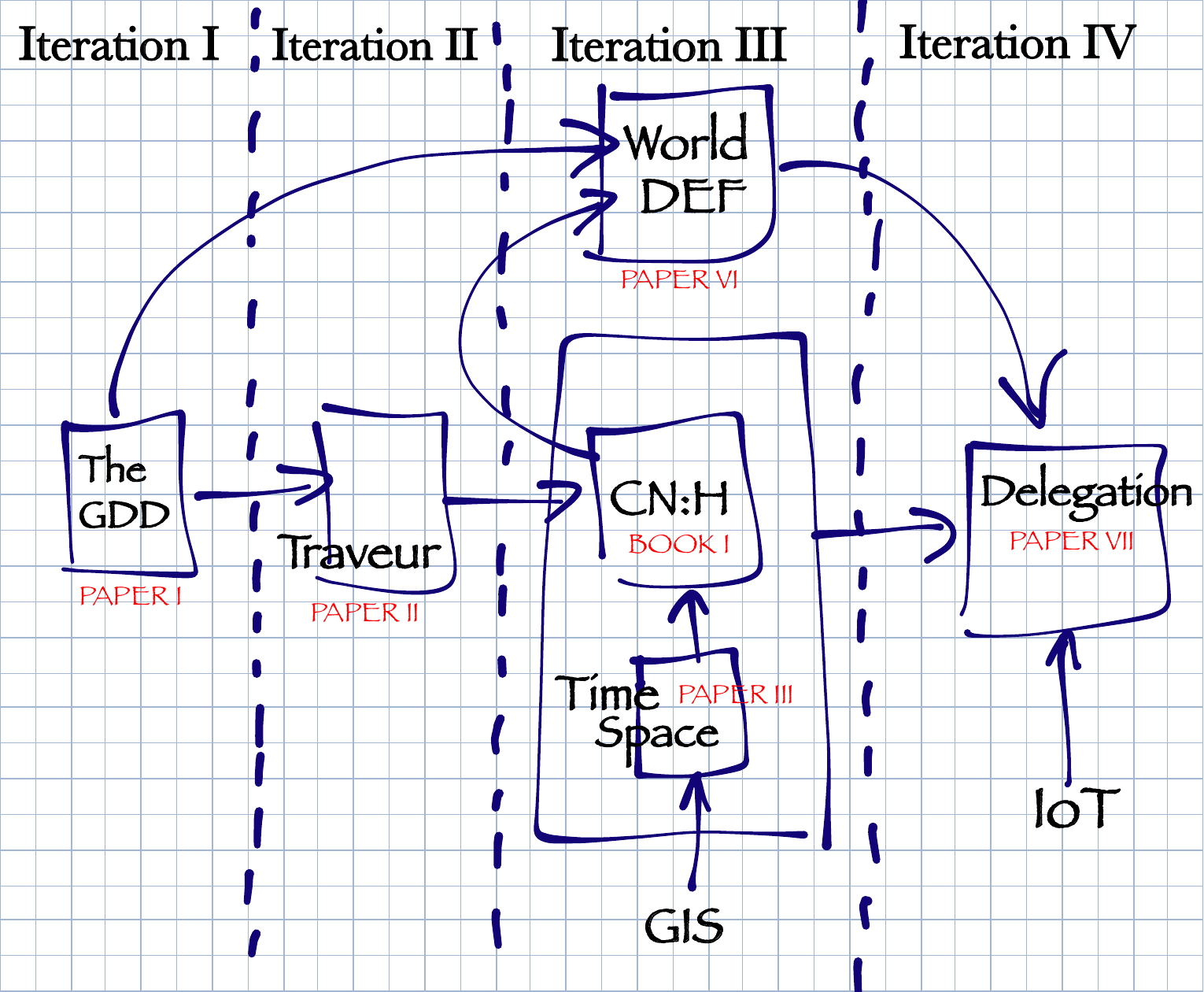}{Dissertation `road map'}{Visual depiction relating publications; organized in columns according to which iteration the publication or project was in, and with arrows showing the flow of influence between projects.
}

The first architecture, discussed in Chapter~\ref{ch:architectures}, is that of the GDD in \mbox{\TheGDD} (see Section~\ref{section:architecture-thegdd}). The GDD proved to model a persistent communication technology, of which the requirements analysis, design and `view' concept influenced the Traveur project. Properties of a persistent communication technology influenced the definition of a `virtual world' in \WorldDEF (marked with the working title WorldDEF in Figure~\ref{figure:dissertation-roadmap}).
Because the architecture of Traveur was never published, a detailed account of the Traveur architecture, and its connection to \Acrobats, is presented in Section~\ref{section:architecture-traveur}. The requirements analysis and design of Traveur influenced \CNH of \mbox{\PervasiveMOO}; poor architectural design lead to broadening the search for technologies to included solutions from the video game industry. Other concepts from Traveur that influenced subsequent iterations were heterogeneous servers, context-awareness and mixed reality. Iteration III is the largest and contains the architecture for \CNH presented in Section~\ref{section:architecture-pervasivemoo}.
\mbox{\PervasiveMOO} contains the requirements analysis for a pervasive games architecture, and a case study evaluating the architecture (see Section~\ref{section:repurposing-vw-engine}). Pervasive games are shown to overlap with virtual worlds, via only a persistence trait. And, that no usable definition for a `virtual world' existed was highlighted. 

\WorldDEF provides a definition, and in Chapter~\ref{ch:combinedwork} it is used to show in detail the overlap between pervasive games and virtual worlds (see Section~\ref{section:vw-defined-overlapping}). Because pervasive games are not required to have Virtual Spatiality, clarification is needed on how to deal with their mixed reality. Using the results from \TimeSpace (marked with the working title TimeSpace in Figure~\ref{figure:dissertation-roadmap}), time and space are mapped from the physical to the virtual, and \viceversa (see Section~\ref{section:mapping-between-virtual-physical}) \ie dealing with the mixed reality of pervasive games. 
In Section~\ref{section:scaling-architectures}, using advances from the domain of Virtual Worlds, all three architectures are compared and scaled up to distributed systems.
\Delegation (marked with the working title Delegation in Figure~\ref{figure:dissertation-roadmap}) explicates the problem of scaling architectures for virtual worlds ({\MMOW}s) and \IoT. In Section~\ref{section:delegation}, results from \Delegation are applied to pervasive games, extending pervasive games into a broader set of pervasive applications incorporating \IoT. This leads to the future work presented in Chapter~\ref{ch:futurework}.

Chapter~\ref{ch:evaluation} validates the research question, verifies the research methodology and evaluates the work in Chapter~\ref{ch:architectures} and \ref{ch:combinedwork}. Chapter~\ref{ch:conclusion} is the conclusion, that summarizes what has been achieved in this dissertation. And, Chapter~\ref{ch:futurework} is the future work section, which discusses the most direct extensions of this dissertation. 
Some of the future work follows directly on \Delegation and Section~\ref{section:delegation}, while some extend from \PervasiveMOO.

%
%

%
%
%

%

%
%

%

\graphicspath{{includes/2_img/}} %

\chapter{Research Strategies / Methods} %
\label{ch:methodology}

Design Science is a special strand of design research, with the \dots
\blockquote{
\dots ``intent to produce and communicate knowledge that is of general interest. [\dots] In contrast [to design], Design Science produces results that are relevant for a global practice \ie a community of local practices, and for the research community. The different purposes of design and Design Science give rise to three additional requirements on Design Science research. Firstly, the purpose of creating new knowledge of general interest requires design science projects to make use of rigorous research methods. Secondly, the knowledge produced has to be related to an already existing knowledge base, in order to ensure that proposed results are both well founded and original. Thirdly, the new results should be communicated to both practitioners and researchers.''\par\hfill --~\citeN[p.8]{johannesson2014-designscience}. 
}
\bigskip

\noindent
``\textit{Design Science} is the scientific study and creation of artefacts [artifacts] as they are developed and used by people with the goal of solving practical problems of general interest''~\cite[p.7, original italics]{johannesson2014-designscience}
``In contrast to empirical research, design research is not content to just describe, explain, and predict. It also wants to change the world, to improve it, and to create new worlds. Design research does this by developing artefacts [artifacts] that can help people fulfil [\sic] their needs, overcome their problems, and grasp new opportunities''~\cite[p.1]{johannesson2014-designscience}. Artifacts are defined here as objects made by humans to address a particular problem; this includes methods and guidelines. Each artifact has an inner structure that can produce certain behaviors \ie offer function for people in a practice. 

As methodological support for projects, Design Science offers researchers the Method Framework for Design Science Research, (see Figure~\ref{figure:designscience-Fig-41}). The method framework consists of five activities that range from problem investigation to demonstration and evaluation; any research strategy can be used in any of the five activities. 
``Many Design Science projects do not undertake all of the five activities of the method framework in depth. Instead, they may focus on one or two of the activities, while the others are treated more lightly''~\cite[p.79]{johannesson2014-designscience}. 
Varied focus is the basis for at least five typical cases of research, namely: problem-focused; requirements-focused; requirements- and development-focused; development- and evaluation-focused; and, evaluation-focused Design Science research. 

\figuremacro{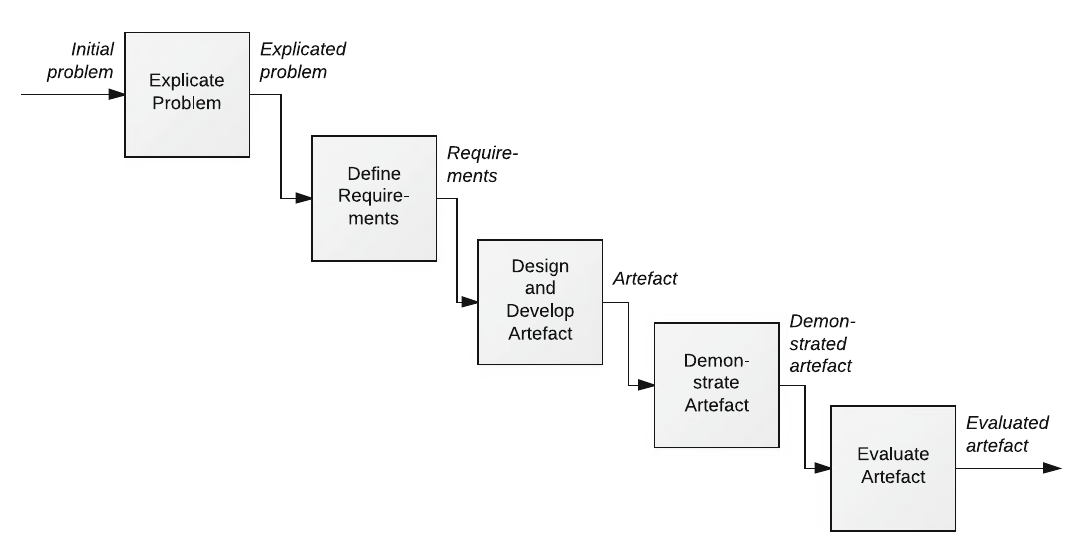}{Method Framework for Design Science Research}{An overview of the five activities in one iteration of the method framework~\cite[Fig~4.1, p.77, with permission of Springer Science+Business Media]{johannesson2014-designscience}; each activity is depicted as a square with input and output on each side.
}

\section{This Dissertation is Design Science Research}

This dissertation is requirements- and development-focused Design Science research~\cite[p.79]{johannesson2014-designscience} for distributed technology-sustained pervasive applications, making use of four iterations (see Figure~\ref{figure:dissertation-roadmap}) of the Method Framework for Design Science research. A pervasive application is a socio-technical system \ie a hybrid system consisting of both artifacts, as well as humans, and the influences that govern their actions. This dissertation is from a technological perspective, but design requirements must still take into account how humans affect the system and are affected. Because the solution maturity in the other domains (\eg Computer Games or GIS) is high and the application domain maturity of pervasive games is low, this dissertation makes use of several exaptations.

\section{Multiple Iterations of Method Framework}

For each iteration of the method framework, a section is included below outlining the focus, the artifact and research strategy used.

\subsection{Design and Development of a New Medium for a GDD}

Iteration I of the method framework was development- and evaluation-focused Design Science research. The artifact for \TheGDD, was a model of a system that could serve as a new medium for a Game Design Document (GDD); a communication tool for game designers, as well as form of documentation. 
Existing GDD mediums were not adequate for the game development community, explicit in previous work by this author~\cite{nevelsteen2008-dev}. Requirements analysis and initial design of the model were performed by this author through document survey \ie reading published criticisms, game development post-mortems~\cite{dingsoyr2005-postmortem} and surveying existing technologies. The model was a `sketch'~\cite{johannesson2014-designscience} of core functionality. 
The design would take much inspiration from existing technologies such as a wiki or the then current technology at the time, called Google Wave~\cite{google2009-wave}. 
Traveur, presented below, was the first project assigned during the PhD studies and the GDD was a side project.
Because development resources were to be spent on Traveur, it was decided that the GDD would not be implemented, but only modeled. 

Design and evaluation of the model was done iteratively in three phases, in collaboration with, and as supervisor for master student, Sergio Gayoso %
\citeN{gayoso2011-gdd}. Formative evaluations at the end of each phase were done with observations in a controlled environment and phenomenological unstructured interviews of participants. As a final summative evaluation of the core functionality exhibited in the model, an interview with designers at a leading game development company was performed and compared to the model.

\subsection{Traveur: Pervasive Architecture and Map Functionality}
\label{section:traveur}

For Iteration II, while others in our research group focused on development and evaluation, this author focused on requirements- and development-focused research.
The initial problem was presented to our research group by the Parkour community stakeholders. 
The artifact for the Traveur project was the design and development of a pervasive game or service for the Parkour community \ie an `instantiation'~\cite[p.29]{johannesson2014-designscience} of a running fully functional prototype, comprised of smaller artifacts (\eg the map function). Artifacts were demonstrated in `local practice'~\cite[p.77]{johannesson2014-designscience} and evaluated as case study. A team of two game designers and four developers would design and development Traveur. ``The project went through four distinguishable phases'': brainstorming workshop, technological probing, prototyping and the implementation of a `fully functional' prototype~\mbox{\citeAcrobats}. 

\subsubsection{Four Distinguishable Phases}

The design goals of the Traveur project were to create a pervasive Parkour game to satisfy the stakeholders and the Parkour community. 
Initial brainstorming was done in a focus group with all stakeholders: the Helsingborg-based Parkour and Freerunning group Air-Wipp; Street Media 7, a small company developing mobile phone technology; and our research team. Air-Wipp was the user representative, Street Media 7 the commercial stakeholder, and our group the research stakeholder.
The collaboration created a wide and diverse scope for the design of the game. 
In subsequent meetings, it became clear that Air-Wipp did not desire any game-like aspects to be developed. First, there was fear that the Parkour community would reject any design that had a competitive aspect (due to the communities' strong emphasis on collaboration), and second, the incitement to interact with people in public, outside the Parkour community, was perceived as very low. The goal of our research group to create a pervasive Parkour game was merged with 
Air-Wipp's goal for creating a Parkour academy function, essentially becoming a training game, that would make it fun to train in a safe way. At the time, no such training game was available, and only towards the end of Traveur did one competing application become available. %
In actuality, Traveur turned out to be a very ambitious and unique project, aiming to combine: mobile community functionality, Parkour training support, a real-time location-based mapping functionality and context-awareness in the form of biometric data. Although the number of active people in the project was adequate, the project was only to last six months. 

\figuremacro{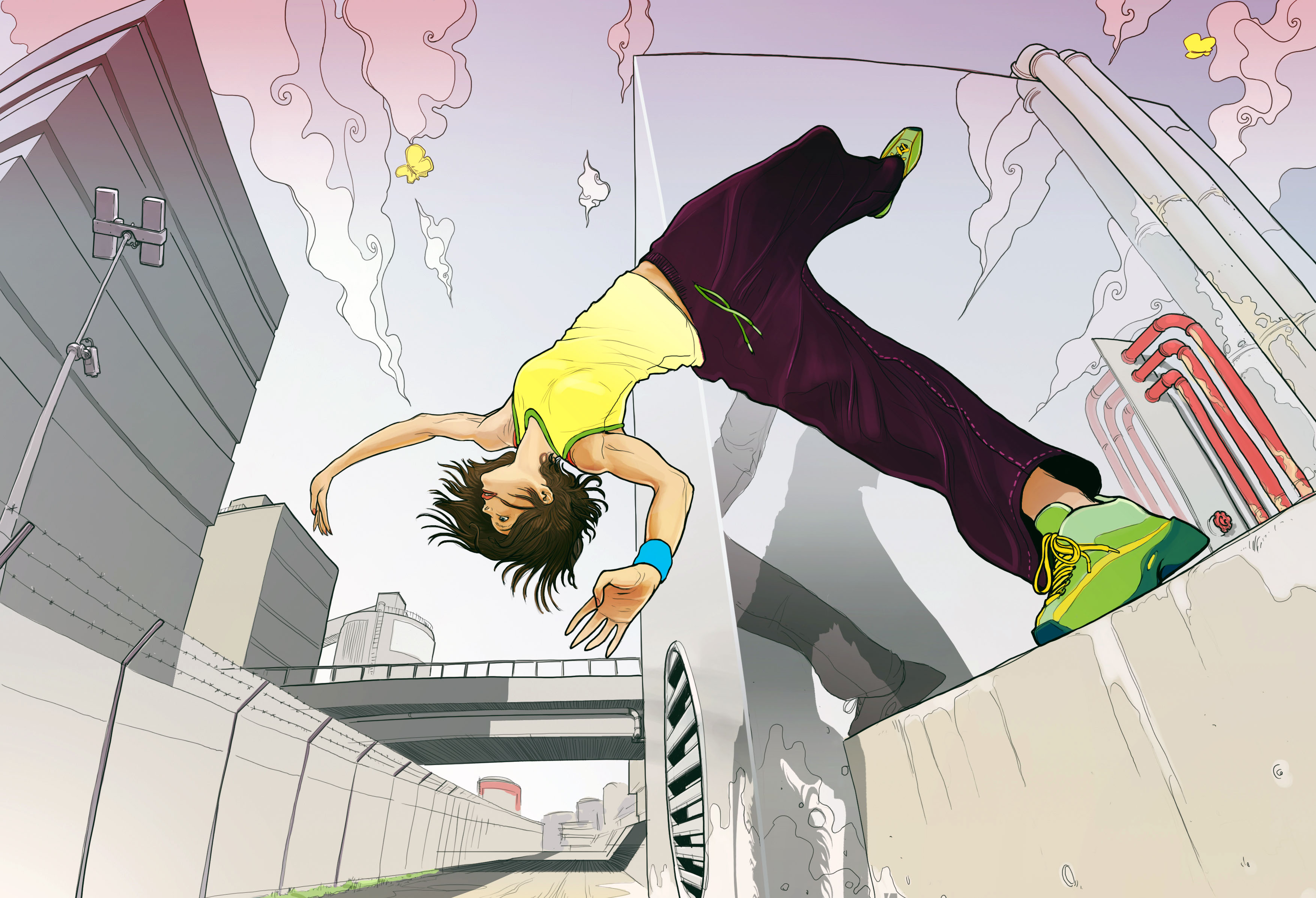}{`Volt'}{Contracted artist, Johan Lindh's rendition of a Parkour vault.
\scriptsize{\url{http://johanart.com/illustrations-StreetMedia7-Traveur.html}}
}

Because Parkour is a high impact sport, it was unclear what technology could be adopted without disturbing their practice. During the second phase of the project, `technology probes' (\ie ``very simple one-function applications that Air-Wipp could use in their daily practice and evaluate for their usefulness''~\citeAcrobats) were built to get early feedback. Extensive work on the technology probes was done by this author, examples of which include: a YouTube~\cite{google2009-youtube} uploader application (app), a simple location-based map with `meet-up' function, and a body harness containing accelerometers. The YouTube uploader was needed, because uploading videos to YouTube via mobile device was not evident or even blocked at the time, to prevent overuse of mobile network bandwidth. The meet-up function, allowed users to see where other users (who were simultaneously using the same meet-up probe) were in real-time. Evaluation of the meet-up probe by Parkour practitioners was considered successful \ie the probe was used as intended, but also `appropriated'~\cite[p.162]{johannesson2014-designscience} as a geo-located chat service and as a tool for playing a game of chase. It was decided to extend the meet-up function and incorporate it into the client prototype (see Section~\ref{section:map_functionality}). 
Since users of Traveur already carried with them a mobile phone, which had a variety of sensors (\eg 3G, GPS, Bluetooth, accelerometers, vibration motor), that enabled context-awareness~\citePervasiveMOO, and Parkour and Freerunning is such a high intensity sport, our team found it interesting to try and capture live biometric data. Prototype harnesses were created that could hold an iPhone in place during the high intensity Parkour jumps and rolls (see an artists rendition of a Parkour vault in Figure~\ref{figure:Traveur-Volt}).
Considering Traveur was designed to be a training guide, it was an advanced feature idea to analyze the accelerometer data, trying to determine if a particular Parkour movement had been accomplished \eg as with \cite{marquez2011-boogies}. Accelerometer data from the iPhone sensors was collected related to speed, impact and rotation, but the data was never tied into the Traveur system. Piggybacking on a health project at the time, biometric data such as galvanic skin response and pulse was also being considered.

Prototyping, in phase three, was done in two iterations~\citeAcrobats; subsequent requirement analysis and design, after each formative evaluation, was done through participatory action research by colleagues. %
Initial design of the architecture was done by Annika Waern, from our research group, and Joel Westerberg, from Street Media 7, with added input from this author. The first evaluation, by Air-Wipp together with their students in Helsingborg, uncovered a need to re-design Traveur, to move it even further away from its original goals of creating a pervasive game. 
The fully functional prototype was again evaluated in a public test with participants from both Stockholm Parkour Academy and Uppsala Parkour. 
An account of how the design of the pervasive Parkour game shifted towards a pervasive service can be found in \mbox{\Acrobats}. A detailed account of the underlying architecture and how it was affected by the shift in the design is presented in Section~\ref{section:architecture-traveur}. 

The fourth and last phase of the Traveur project focused on `value conflicts uncovered' during the previous tests. Traveur was evaluated again and subsequently redesign, resulting in \Acrobats.

\subsubsection{Map Functionality}
\label{section:map_functionality}

Initially, development of the mobile client and the Log-of-Everything were assigned to this author. Because Traveur turned out to be a highly ambitious project, a consultant was hired to develop the client instead, alleviating this author to implement the technology probes and the map functionality; design of the map functionality was done by this author, under supervision of Annika Waern. Initial map functionality grew out of an idea from this author, to use a service such as Find My Friends~\cite{apple2011-findfriends} or Google Latitude~\cite{google2009-latitude}, to allow Parkour practitioners to find each other. At the time, such location sharing applications were only then becoming widely accepted.
A map function was created which allowed a number of users to temporally share their location with others in real-time and leave geo-located comments on the shared map. 

Considering the Traveur interface was already offering users a map, it was decided to extend the map functionality~\citeAcrobats (see Figure~\ref{figure:Traveur-IMG-9597}) to a location-based mobile platform supporting: areas of interest, training data, various media types (text, sound, image and video), historical traces, a `heat map'~\cite['Heat map']{wikipedia.com2015} and a meet-up function. 
This lead to the following list of requirements:
\begin{compactitem}
\item a shared map showing areas of interest with various levels of detail;
\item real-time response \ie updating of data every second;
\item robust handling of mobile network and GPS `uncertainty'~\citePervasiveMOO;
\newpage
\item resolving the trade-off between coarse GPS updates (\ie not suitable for tracking walking or Parkour running) and detailed GPS updates (\ie battery life and not being able to use the service in a moving vehicle).
\item ability to add content to a shared map (\ie crowd-sourcing of content), with new content updated on all clients (\eg in a heat map of activity); 
\item and, the display of historic data on the map \eg showing movement traces over time.
\end{compactitem}
Areas of interest would be geo-located points or polygonal areas of ongoing events, or good or bad training locations. To each area of interest training data could be assigned showing what Parkour exercises could be performed there. By allowing for various media types to be assigned to areas of interest, location-based content could be crowd-sourced from the Parkour community. 
Historical data on the map would allow for traces of movement over time to be shown on the map; this would allow Parkour runners to see a trajectory through the city that others might have taken previously. And, a heat map showing areas with a high level of activity was designed for, but never implemented. 

\bigskip
\figuremacro{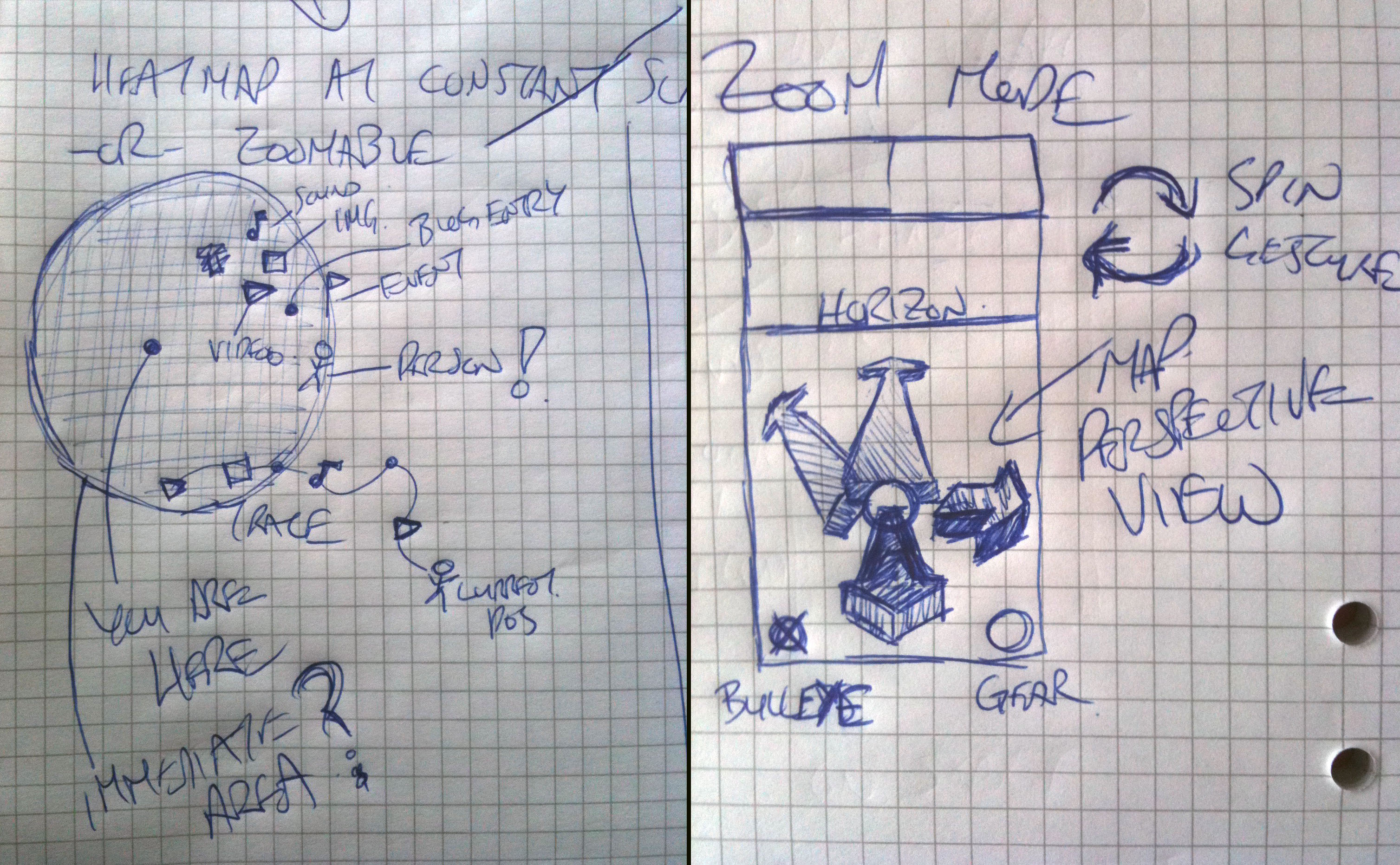}{Two images of the map functionality napkin design}{on the left, a circle of interest with various content contained in it, and on the right, an interface sketch of a perspective view, with spin gestures and buttons.}

\subsection{Pervasive Games Architecture: Pervasive MOO}

Iteration III was largest, with a focus on requirements, development and the evaluation of an architecture for a pervasive game, called Codename: Heroes.
Requirements for \CNH were provided informally as a ``long term pervasive game'', spanning months or years, and developed `in house'. A single game designer and two developers would design and develop \CNH under the supervision of Annika Waern, who also contended with design. The artifact being an instantiation of a system capable of staging a pervasive game. The artifact is the same for the following subsections \ie publications \mbox{\PervasiveMOO}, \mbox{\TimeSpace} and \mbox{\WorldDEF}. But, since the focus and research strategy differed for each study, they are presented in each of the sections below.

\subsubsection{Survey of Pervasive Games and Technologies}

Given the limited resources available to produce \CNH, the problem made explicit by the Traveur project (see Section~\ref{section:architecture-traveur}) was that if \CNH was to be successful, an architecture was needed that could cater to the requirements of a pervasive game. In order to find a feature set describing a would-be pervasive games engine, a systematic review~\cite{ampatzoglou2010-gamesreview} was performed as a survey of existing pervasive game projects and technologies. 
Chapter~2 of \PervasiveMOO is that systematic review, serving as an informal requirements analysis for a would-be pervasive games engine.

\subsubsection{Virtual World Engine Staging a Pervasive~Game}

The focus for Chapter~3 of \mbox{\PervasiveMOO} was development and evaluation; the feature set discussed above was implemented in a running system and evaluated for feasibility \ie a case study was the chosen research strategy.
Using the feature set from above, a virtual world engine was chosen as an architecture in the same `product line'~\cite{bass2013-inpractice} as a would-be pervasive games engine~\citePervasiveMOO; for the implementation of \CNH, a virtual world engine could be exapted to stage a pervasive game \ie a solution from the domain of Computer Games exapted to the domain of Pervasive Games. 
The \CNH architecture was demonstrated twice in two separate stagings of the game. Evaluation of \CNH from a cultural perspective was done by colleagues.

\subsubsection{Exaptation of the Triad Framework}

A problem that was made explicit in the case study of \CNH, was the complexity of dealing with mixed reality \ie mapping the virtual to physical, and \viceversa. A model dealing with such mappings represented a small artifact in the larger \CNH project. Rather than develop a new model and because both domains are based on the Earth's geography, \TimeSpace presents an exaptation of the Triad Representational Framework from the domain of GIS to Pervasive Games. The focus of \TimeSpace is evaluation research verifying the model to be applicable to pervasive games, using simulation \ie successful implementation. 

\subsubsection{A Definition for `Virtual World'}

Another problem that was made explicit in the case study of \CNH, was that a working definition for `virtual world' was lacking. Virtual world engines existed, but since no working definition of `virtual world' existed, there was no list of properties an engine had to exhibit in order to be called a `virtual world engine'. The focus of \WorldDEF was to determine the requirements for an artifact construct in the form of a definition \ie the requirements for a virtual environment to be called a `virtual world' by definition. The definition represented a small artifact construct, which could be used in combination with the virtual world engine, in the larger \CNH project. Because no consensus has been reached on properties of a virtual world, that coincided with technology implementing a virtual world, the definition was obtained using grounded theory rather than a literature review. Theoretical sampling was used to select technologies that refined the definition and this was continued until theoretical saturation was achieved (see Section~4 of \mbox{\WorldDEF} for details). To evaluate the resulting properties, `discriminant sampling'~\cite{creswell2013-qualitative} was used; and, the resulting definition was evaluated by comparing it with prominent existing definitions. 
The effectiveness of the obtained definition was demonstrated by creating an ontology of virtual worlds, and classifying advanced technologies, such as: a pseudo-persistent video game, a MANet, virtual reality, mixed reality, and the Metaverse. 

\subsection{Aligning Research of {\MMOW}s with that of \IoT }

The focus of Iteration IV is problem-focused Design Science research, explicating the problem of designing and developing an architecture for distributed pervasive applications. Properties, related to the scaling of architectures, are gathered from the domain of {\MMOW}s and \IoT through a document survey; properties found in the domain of {\MMOW}s are then evaluated against those found in domain of \IoT. The artifact here is a distributed technology-sustained pervasive applications, which can be designed and developed in future research (see Chapter~\ref{ch:futurework}).

\graphicspath{{1_relatedwork/figures/}} %

\chapter{Related Work} %
\label{ch:relatedwork}

In \PervasiveMOO, a systematic review was performed to find the then current state of the art in pervasive game architectures. Several publications were cited suggesting frameworks and middleware for pervasive games (\ie custom-built solutions, see Section~2.4.1 of the book), but the solutions are not compared with existing game engine technology (more `general-purpose' solutions \cite{gregory2009}), and only a few publications repurpose an existing game engine in their project \eg Ambient Wood~\cite{thompson2003} or ARQuake~\cite{thomas2000-arquake}. If that review is extended to the day of this writing, then the conclusion remains the same \ie still no wide-spread use of advances in existing video game engine technology for pervasive games. 
\citeN{viana2014-devreview} present a systematic review of research on software engineering pervasive games; their study shows an increase in development in the area of pervasive games in general, but with areas such as "Reuse" and "Interoperability" being underexplored. Their review is said to include three papers on software system architectures (\ie [S37, S74, S83]) that, similar to other recent publications \cite{fischer2014-mapattack} and \cite{pimenta2014-ubigameengine}, provide custom-built solutions, but do not compare with the large body of knowledge found in the area of video game engines. \citeN{pimenta2014-ubigameengine} recognize the need for a pervasive games engine, but only relate to technologies in the domain of Pervasive Games (\eg MUPE). If pervasive games, do not take into account the domain of Computer Video Games, then developers of pervasive games will be left continually reinventing existing technologies. 

\citeN{paelke2008-lbgaming} state that many pervasive games ``implement their own custom game engine by tying together available standard components for distributed applications and filling the gaps with custom code''. They demonstrate that a web server can serve as a game engine when tied to other components. Although a viable solution, ``tying together available standard components'' does not provide any foresight into how the architecture will behave with respect to scalability. If the game is suddenly successful, how must the architecture change to handle an increased load due to the influx of additional users? Also, after repeatedly creating many solutions for various pervasive games, one might want to collect commonly used components into an engine \ie find out which characteristics are important for pervasive games and build an engine to accommodate them. A recent publication, by \citeN{valente2015-requirements}, does this by providing a formal requirements analysis for pervasive mobile games.

ARQuake~\cite{thomas2000-arquake} is a version of the original game called Quake~\cite{id1996-quake}, extended into a pervasive game. ARQuake was developed by repurposing the original Quake engine, as mentioned by \citeN{lewis2002}. 
Quake and ARQuake are similar to Doom (presented in \mbox{\WorldDEF}) in the sense that they both do not have a persistent virtual world.  Because a persistent virtual world was not required by the game, ARQuake could be implemented by repurposing a ``graphics engine rather than a virtual world engine, with a Euclidean spatial representation, that could be directly overlaid onto the physical world''~\citePervasiveMOO. Without a persistent virtual world, ARQuake's times of play (when the player is in the game world) can not be temporally expanded~\citePervasiveMOO \ie ``uncertain, ambiguous, and hard to define''~\cite{montola2009}.
A recent survey, by \citeN{kasapakis2015-trends}, aims to classify pervasive games into age generations based on technologies used. The survey is limited to 18 pervasive games \eg not containing tabletop games, smart toys and trans-reality games, even though being mentioned as sub-genres in the articles' related work. 
\citeN{kasapakis2015-trends} state that ``the game engine organization model is largely dictated by the game scenario to be supported'' and this author agrees. If a general-purpose pervasive games engine is to be created, commonalities between game technologies must be found \eg support for a virtual persistent world. 
Each engine, in the domain of Computer Video Games, has advantages and disadvantages which must be considered when repurposing it as a would-be pervasive games engine \eg the focus of a graphics engine is typically high definition graphics for a few players, with less focus on building a large expansive world. 
\citeN{kasapakis2015-trends} continues that the ``current technological status favours [\sic] always-on connectivity, hence, centralized models''; this is not entirely correct since, ``always-on connectivity'' relates to persistence, which can also be obtained through decentralized models.

In the Ambient Wood~\cite{thompson2003} a MUD-based virtual world engine (with support for persistence) was chosen to stage the game, in order to ``to replicate and model the physical world so that interactions between devices in the physical environment would have corresponding interactions within the model''~\mbox{\citePervasiveMOO}. Ambient Wood was a client/server architecture, with a custom-built proxy object called MEAP~\citePervasiveMOO handling the various heterogeneous devices \eg mobile handhelds as sensors or sound actuators. Although MEAP was custom-built, it was developed as a more general-purpose tool for heterogeneous devices~\cite{see2001}. There is no mention in the literature of possibly reusing the architecture of Ambient Wood to stage additional pervasive games, or if the architecture could be scaled to support more players.
The architecture of Ambient Wood was a distributed system of heterogeneous devices, with one centralized MUD game engine. %
\citeN{thompson2003} stated that they were extending their work across distributed MUD engines in \citeyear{thompson2003}, with an interest in the ``delegation of ownership [authority] to localised [\sic] MUDs''.

In a recent publication, \citeN{laine2015-distributedexer} mention the novelty of the pervasive game (exergame) Othello as being distributed gameplay, enabling ``players from anywhere in the world to play against each other''. The architecture, however, was client/server ``causing a noticeable delay [latency]''  in communication to due the geographical distance between client and server. \citeN{laine2015-distributedexer} note that ``unlike MMORPGs [{\MMOW}s with RPG theme], it is not necessary to divide computing across multiple servers [for Othello] because there are few concurrent players''. %
In a recent publication by \citeN{kamarainen2014-cloudgaming}, the subject of cloud computing is discussed as a possible solution for pervasive and mobile computing, allowing the ``end-user device to offload computation, storage, and the tasks of graphic rendering to the cloud''. Similar to Othello, \citeN{kamarainen2014-cloudgaming} remark that latency is the ``main challenge'' for cloud gaming, with most interactive games requiring response times that only ``local deployment scenarios'' can deliver. As a solution, they ``propose to use [a] hybrid and decentralized cloud computing infrastructure, which enables deploying game servers on hosts with close proximity to the mobile clients''. To exploit local resources, the Fun in Numbers (\FiiN) platform (presented in \PervasiveMOO) features a distributed multi-tiered (\ie four layered) large-scale architecture~\cite{chatzigiannakis2011-implement}. The \FiiN architecture supports more than one game engine, with each engine being the ``local authority for each physical game site''~\cite{akribopoulos2009-sensornets}. All game engines are coordinated by a centralized top-most layer. The bottom layers of the \FiiN architecture enable support for ad-hoc networks and \IoT. The problem with the \FiiN architecture is that it is unclear exactly what types of pervasive games are supported. Pervasive games are defined in the \FiiN publications as ``games played in the physical space, indoors or outdoors, using mobile handheld devices, context-awareness, and in certain cases some degree of infrastructure and scripting''. %
\citeN{chatzigiannakis2011-implement}  make no distinction between technology-sustained or technology-supported pervasive games, even though technology-supported pervasive games do not necessarily require a game engine %
\ie the infrastructure to enable them is very different. \citeN[original italics]{akribopoulos2009-sensornets} state that \FiiN targets ``mainly games that \textit{involve multiple players, rapid physical activity, gesturing,} \dots~and \textit{less storytelling-based games}'', which could account for why game master interfaces are not present in the architecture~\mbox{\citePervasiveMOO}.

Above, it is stated that a web server can function as a pervasive games engine. %
In an early publication, \citeN{broll2009-perci} describe the Perci Framework, which connects mobile devices (via a proxy) to a web server, forming a pervasive service that leverages \IoT \eg mapping the virtual to the physical through digital devices. The web server provides a persistent virtual world and it is the ``nature of persistence [that] is closely related to the nature of ubiquitous as understood in the Ubiquitous Computing research field''~\cite{lankoski2004-songsnorth}. The Perci Framework can be used by games or other applications, and is in the domain of Ubiquitous Computing. 

%

%
%
%

%

%
%

%

\graphicspath{{includes/4_img/}} %

\chapter{Design of Three Software System Architectures} %
\label{ch:architectures}

From the year 2000 to around 2009, there was lots of activity in the research community to build pervasive games, using either custom-built solutions or now outdated engines~\citePervasiveMOO \eg MUD or MUPE. Without the advantage of hindsight, creators of pervasive games could not generalize the requirements for pervasive games. Considering video games have existed for decades, with reusable game engines to drive them, this dissertation sets out to see if engine technology from the domain of Computer Video Games can be repurposed to stage technology-sustained pervasive games. 
Discovering this was done in practice through the design of three software system architectures:
the~GDD~\mbox{\citeTheGDD}, 
Traveur~\citeAcrobats, and
\CNH~\citePervasiveMOO.

This dissertation uses four iterations (one per architecture mentioned above and one presented in Chapter~\ref{ch:combinedwork}) of the Method Framework for Design Science Research~\cite{johannesson2014-designscience} to obtain requirements for distributed pervasive applications: 
Iteration I was a design exercise on how to model a persistent communication technology (the GDD), of which the requirements analysis, design and `view' concept, influenced subsequent iterations.
The requirements analysis and design of a pervasive service (Traveur) from Iteration II influenced Iteration III, leading to a search for architecture solutions from the video game industry, and the influence of concepts such as heterogeneous servers, context-awareness and mixed reality.
Iteration III provides a characteristic feature set for pervasive games (evaluated through \CNH), aligns pervasive games research with that of virtual worlds (based on the persistence trait), shows how the Triad Representational Framework can be applied to pervasive games~\citeTimeSpace, and a definition for `virtual world'~\mbox{\citeWorldDEF}. 

\section{A Persistent Communication Technology}
\label{section:architecture-thegdd}

The GDD of \TheGDD is an application with already some pervasive characteristics \eg temporal expansion. The requirement analysis for the GDD resulted in four communication-based requirements for the medium:
\begin{compactitem}
\item collaborative user editing with enabling communication mechanisms;
\item being readily available at all times to all users \eg web based;
\item changes are communicated to the users, with differentials; and, 
\item support for a variety of different discussion channels \eg real-time and non-real-time based on video, audio or text.
\end{compactitem}
And seven additional general requirements:
\begin{compactitem}
\item a mechanism to allow for narrative linearity and linear printing;
\item editor roles with an option to force edits to be approved by a lead editor;
\item a familiar user-interface and intuitive interaction model;
\item quick updating, with fast access and editing;
\item many media/file types \eg audio, video, images, spreadsheets, \textit{et cetera};
\item the ability to link relative information, with auto-linking;
\item and revision control tracking and a backup system.
\end{compactitem}
Many of these requirements were common and in wide-spread use at the end of 2011, when the GDD project was started, and others not so common. Google Wave~\cite{google2009-wave} had already been current for two years and Google Docs~\cite{google2009-docs} was available, but not yet a mainstream service open to the public \ie collaborative user editing via the Internet was not mainstream, as it is at the time of this writing. 
An important finding in the GDD work, which is still not yet in wide-spread use as of this writing, is the notion of a `view' \ie a subset of the available data as a customized visualization for a particular user~\citeTheGDD. 

Using \WorldDEF, the modeled GDD medium can be categorized at least as a persistent communication technology (see Section~\ref{section:gdd-vs-virtual-world}) \ie a persistent environment, spanning one shared data space, where users can interact in real-time. And, it is these properties that influenced both Traveur and \mbox{\PervasiveMOO}. In essence, the GDD served as an exploratory case study for pervasive applications.

\section{Design of a Pervasive Parkour Game;\\ shifting requirements and mashed-up components}
\label{section:architecture-traveur}

%

%
%
%
%

The second architectural case study was that of Traveur, which had major architectural differences in design when compared with the previous project, the GDD. In Traveur, participatory action research was to blame for fluid project requirements, highlighting that, in a pervasive setting ``heterogeneity is the norm''\footnote{It was Annika Waern who coined this phrase.} \ie mashed-up technologies with distributed computing. 
In comparison to the GDD, the GDD medium was designed to `readily available', but it applied only to those who were already behind a computer desk designing or developing a video game, not to those with mobile devices out in the physical world. In contrast Traveur would strive to be a pervasive service for the activity of Parkour running, with the mobile phone as an interface to a virtual world \ie striving towards ubiquitous availability and ubiquity of access~\citePervasiveMOO.
Out of the architectures studied in this dissertation, Traveur was the most context-aware, making use of a wide variety of location dependent and biometric data. This data was captured through sensors located on the runners and could be displayed through the map functionality.

\subsection{Architectural Back-end}

The initial architecture was done by Annika Waern, from our research group, and Joel Westerberg, from Street Media 7, with added input from this author. 
The architectural discussion was straightforward until a decision was needed on where persistent data was to be stored. Since Street Media was building a more general community platform based on their proprietary back-end, `Streetside', it was argued that Street Media needed to maintain control over the community data. However, the existing `Creator' component, from our research group, which was to handle the game aspects of Traveur, already had its own database connectivity. Both Streetside and Creator had been created to be a stand-alone central components in a system \ie neither had built-in functionality to replicate data to other servers. At this point, the decision was made to make Traveur a `mash-up'~\cite{singhal2013-multicloud} of existing services into one application, allowing both parties control over their segment of the data.  
Because neither of the two systems would have a complete set of the data, a component called the Log-of-Everything\footnote{It was at this point that the architectural sketch of the back-end, with the redirection of data and the Log-of-Everything, raised an eye-brow of this author.}
was to be created such that it aggregated the data from Streetside and Creator (see Figure~\ref{figure:Traveur-log-of-everything} for the original design on whiteboard, and Figure~\ref{figure:Traveur-log-of-everything-sketch} for a sketch thereof). Aggregated data could then be exported to other services. %
Due to lack of resources, the Log-of-Everything was never created.

\figuremacro{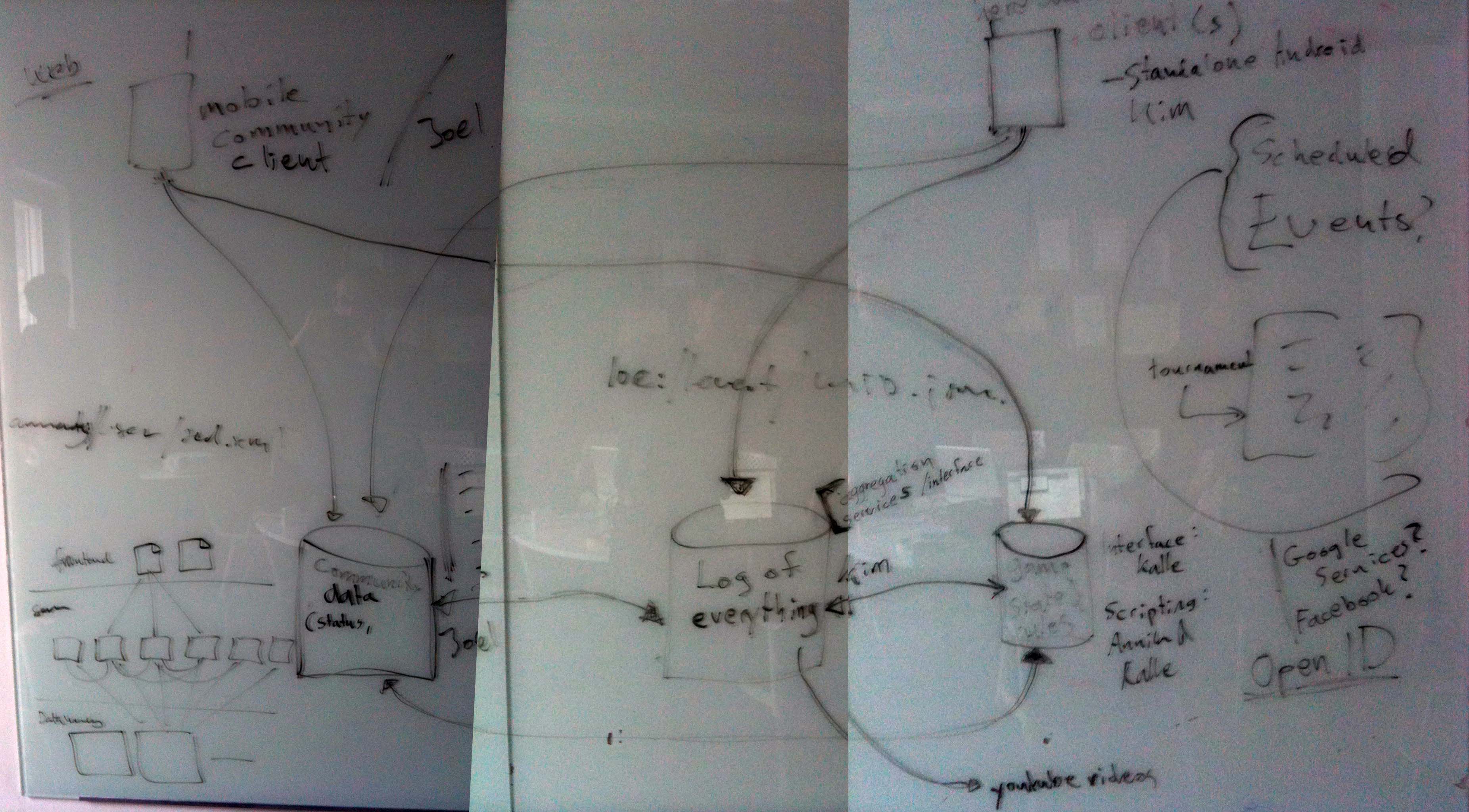}{Traveur architecture}{Initial design on whiteboard; the top two squares represent one or more client applications, each cylinder represents a database of information, and arrows show the flow of information (for clarity see the sketch in Figure~\ref{figure:Traveur-log-of-everything-sketch}).}
\figuremacro{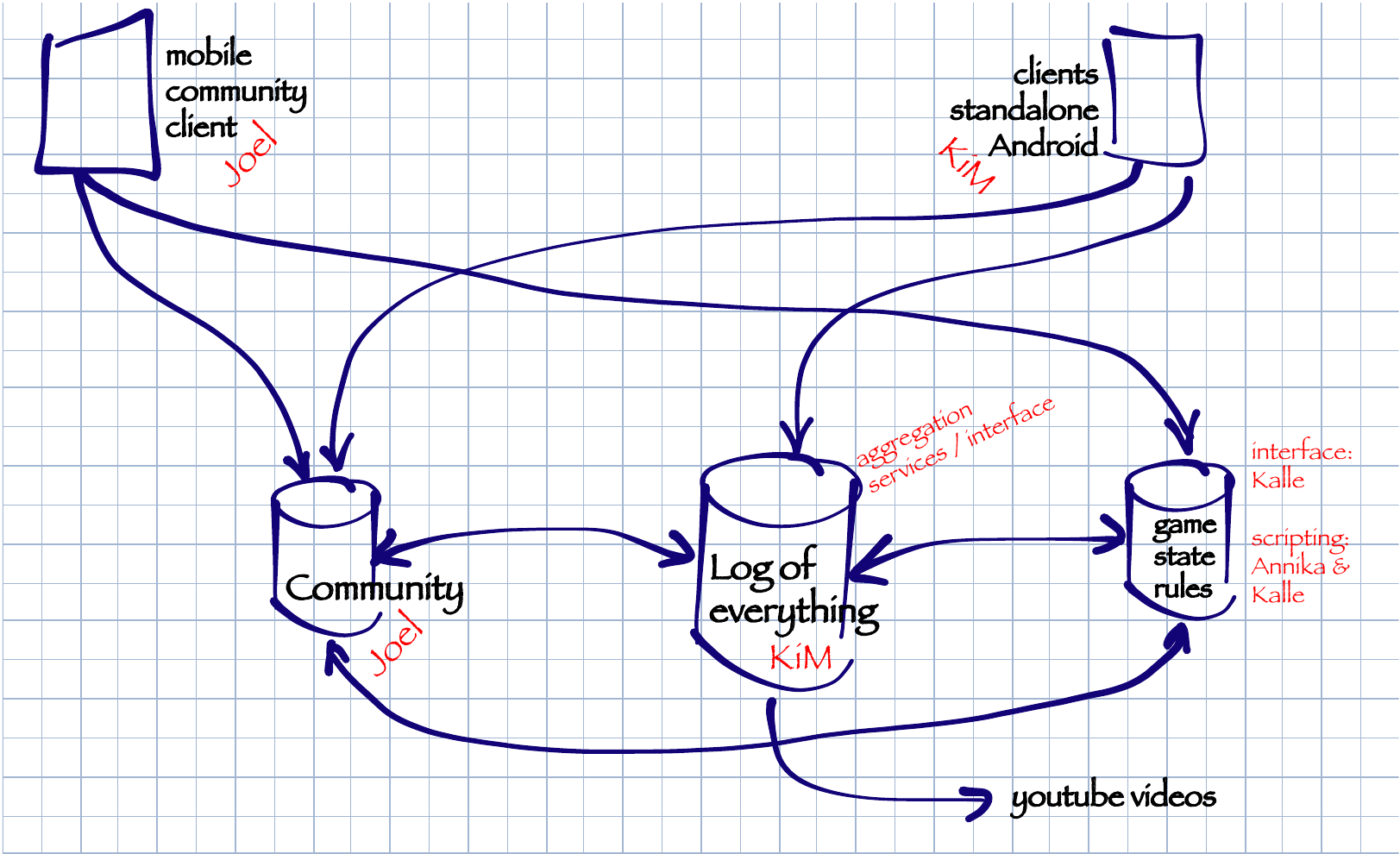}{Sketch of the Traveur architecture}{See Figure~\ref{figure:Traveur-log-of-everything} for a picture of the original whiteboard and a legend. The text in red are notes labeling who is responsible for what component \eg for game state and rules component, interface was to be handled by Kalle, and scripting by both Annika and Kalle.}

\subsection{Architectural Front-end}

Initial prototyping was done on Android~\cite{google2008-android} based mobile phones, but one of the first shifts of the design requirements meant moving to the iOS~\cite{apple2007-ios} mobile platform. Android was initially chosen because it was an open platform, allowing for direct access to phone hardware and an easier application publishing procedure (Apple employs an approval process through which all apps are screened before being published, making the distribution of prototypes difficult). But, feedback came from the Parkour community that most practitioners where using iPhones running iOS; design of the mobile client for Traveur project became that of an iPhone app using the native development framework to access the phone's hardware and sensors. The user interface consisted of five different `tabs': 
\begin{compactitem}
\item News, a rolling blog of all news and events in the community; 
\item People, simply a long list of all participants in the system; 
\item Skills, which represented the game functionality provided by Creator; 
\item Map, an augmented WebMap implementing the map functionality (see Section~\ref{section:map_functionality}); and, lastly a \item User Profile, to control user login. 
\end{compactitem}
Initially, News, People and Map tabs were filled using data from Streetside, leaving the Skills tab to be filled by Creator. The User Profile tab did not exhibit proper functionality until later in development. Because content was being supplied by two different servers, this meant that a connection to each server needed to be maintained \ie lack of connectivity to a server meant that those tabs the server was responsible for were blank.
Because the set of user profiles logged in the community function was larger than that of the game (since login by the general public into the community, not participating in the game, was considered ok), authentication of user identity (\ie control over the logins) was delegated to Streetside, and deemed to be replicated to Creator.
This changed the dependency graph so that Streetside handled all client communication and proxied data for Creator (see Figure~\ref{figure:Traveur-dependency-graph}) \ie Streetside could stand alone, but Creator was dependent on Streetside to replicate user data. Streetside was dependent on Creator, but this dependency was actually the client's dependency passed through Streetside \eg if Creator failed, Streetside would simply not be able to provide the client with the required data, but Streetside would not suffer from it. Because of the chosen replication algorithm, the order in which the services were started and the stability of each machine became an issue.%

\figuremacro{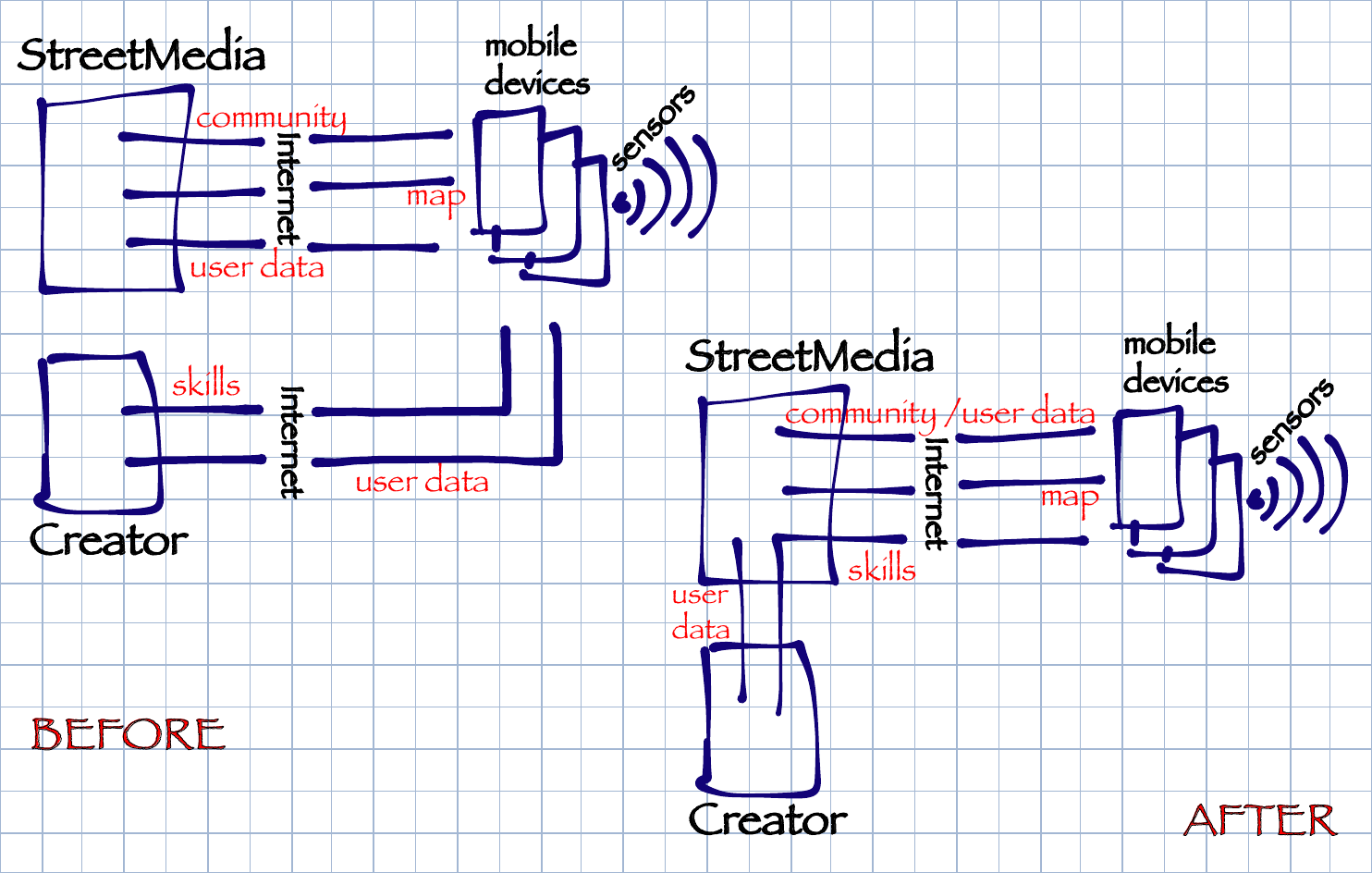}{Traveur's change of the dependency graph}{On the left (BEFORE): StreetMedia and Creator servers were standalone, with no user data shared between them. On the right (AFTER): all mobile client communication is with StreetMedia and user data is replicated between the two servers.}

\subsection{Heterogeneity is the Norm}

Because Traveur aimed to develop a service under fluid project requirements and on top of existing heterogeneous technology, problems could have been anticipated. Neither Streetside nor Creator was constructed to work together with another system \eg both systems had their own database and internally assumed they had a complete `primary copy'~\cite{yahyavi2013-p2pmmog} of the data.
Synchronization of the two systems, to ensure consistency of replicated user data, was implemented poorly, resulting in large amounts of latency. 
Because two stakeholders wanted control over the data, the dependency graph was such that the phone client depended on two dedicated servers, with an additional dependency between them (see Figure~\ref{figure:Traveur-dependency-graph}, AFTER). These dependencies were particularly fragile in a mobile setting where `uncertainty'~\mbox{\citePervasiveMOO} in network connectivity is problematic.
Creator supported scripting, which was useful to change the rules of the game on the fly, but this usefulness was limited to the Skills tab \ie did not extend to Streetside or the client. In order to support context-awareness using biometric data (\eg galvanic skin response), the architecture would have to support streaming data, but neither Streetside or Creator were built with streaming in mind.

\newpage
Essentially, Traveur fed into the next project the notion of how \textbf{not} to do the architectural design for a pervasive system. 
The next architecture to be built, for \CNH~\citePervasiveMOO, was to have a significantly smaller development team and less resources, so choosing the appropriate technologies for it would be critical. This lead to broadening the search for technologies to include solutions from the video game industry \ie not waste time implementing critical functionality that already exists in other technologies. Also from Traveur, it was obvious that, since neither Streetside nor Creator were `mature'~\citePervasiveMOO systems, it had crippled their reliability, and if an architecture for a pervasive game was to provide ubiquitous availability, it would have to be reliable~\citePervasiveMOO.

\section{Pervasive Game Architecture}
\label{section:architecture-pervasivemoo}
\label{section:repurposing-vw-engine}

For \CNH, this author assumed the role of systems architect.
In order to avoid the mistakes made in Traveur and to be able to choose the appropriate technologies for \CNH, a good understanding of the requirements for a pervasive games engine would be needed. 
Although there are plenty of publications on pervasive games, it would seem the majority are from a cultural perspective, designing the game, rather than from a technological perspective, describing the technical requirements~\citePervasiveMOO. Many pervasive game projects have been undertaken in the years of Equator and IPerG, but literature lacks describing common overlapping characteristics of the technologies needed for pervasive games. This is understandable, because during the Equator and IPerG years, architects of those projects could not have effectively determined generalized properties of a pervasive games engine, because they did not have the advantage of hindsight \ie it is far easier to determine requirements for an architecture in hindsight, rather than when faced with the problem~\cite{bass2013-inpractice}. 
\citeN{brooks1995} refers to this as ``planning to throw one away; you will, anyhow''.
That a literary survey was needed, seems to be corroborated by the fact that smaller similar surveys %
were being performed around the same time (see Section~2.4.1 of \PervasiveMOO).
Chapter~2 of \mbox{\PervasiveMOO} is a systematic review into characteristic engine features that describe a \mbox{would-be} pervasive games engine.
These features can be considered a set of informal requirements from which a set of formal requirements can be drawn. 
The assumption is made that, if a would-be pervasive games engine is to be general-purpose it should have support for the found characteristics, even though not all pervasive games will make use of them \eg World Persistence, Game Master Intervention, and Reconfiguration, Authoring and Scripting in Run-Time.

\newpage
\noindent
In the survey of \PervasiveMOO, the following component feature set was obtained: 

\begin{description}
\item[Virtual Game World with World Persistence]: a spatiotemporal instance, with interacting virtual elements (at least one of which being the player); a game world that overlaps with both the virtual and the physical world; a world that continues to exist and develop internally even when there are no people interacting with it (persistence); and, ubiquitous availability
through a reliable architecture.

\item[Shared Data Space(s) with Data Persistence]: a common shared data space, with coordinated communication to it; and data persistence in the event of a shutdown or system failure \ie fault tolerant and recoverable.

\item[Heterogeneous Devices and Systems]: support for non-standard input devices, comprised of sensors and actuators, that form an interface between the player and the game; resolution of interoperability issues through a device abstraction layer; and the use of service-oriented architectures, or the offering of such services.

\item[Context-Awareness]: context information \eg location, body orientation, available resources including network connectivity, proximity to surroundings or noise levels; context information is obtained through sensor enabled heterogeneous devices or service-oriented architectures; dealing with uncertainty in position localization or networking.

\item[Roles, Groups, Hierarchies, Permissions]: various roles for player and non-player characters, organized in groups or hierarchies; different permissions or privileged information for the various roles or organizations, perhaps through an entirely different interface to the game.

\item[Current and Historical Game State]: including semi-static player info; a view of the current internal game state \eg through direct inspectable properties or through a specialized management interface; a historical perspective of the game state \eg through the logging of event data for post-game analysis; or any meta-level game information, such as game master documentation.

\item[Game Master Intervention]: the semi-automatic execution of the game through game master intervention in run-time \eg by directly manipulating the internal game state or through specialized interfaces, that potentially translate massive amounts collected game data into a human consumable form; game master intervention can be provide by a service-oriented architecture.

\item[Reconfiguration, Authoring and Scripting in Run-Time]: in the pre-game phase (\eg for location adaptability), that can be extended into the in-game phase; data-driven reconfiguration of software, hardware and devices; dynamic story and content through content generation in-game; changing of the game rules through data-definition or run-time languages; autonomous agents; or the simulation of events for the Wizard of Oz~\cite{dow2005-mrwoz} technique (see Section~2.3.8 of \PervasiveMOO).

\item[Bidirectional Diegetic\footnotemark~and Non-Diegetic Communication]:\footnotetext{``The `diegesis' of a story consists of whatever is true \textit{in that story}. Diegetic elements are `in the story'; non-diegetic elements are not.''~\cite[original italics]{bergstrom2011-framing}} through various channels and/or interfaces.

\end{description}
\noindent

Many of these features are quite generic (\eg Current and Historical Game State) and so are supported by engines in the domain of Computer Video Games. Features more specific to pervasive games are Heterogeneous Devices and Systems; Context-Awareness; Game Master Intervention; and Reconfiguration, Authoring and Scripting in Run-Time; and Bidirectional Diegetic and Non-Diegetic Communication \eg non-diegetic communication is required in a virtual world, but not as pronounced as in a pervasive game where it is needed to cope with social expansion. 

\begin{figure}[htbp]
	\centering
	\includegraphics[width=.9\textwidth]{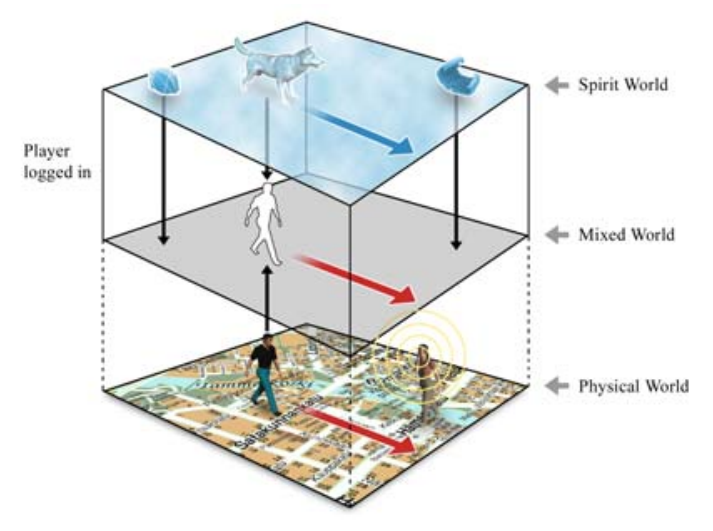}
	\caption[Coextensive worlds]{\textbf{Coextensive worlds} - A depiction of the virtual (Spirit) world coextensive with the physical world; when the player logs in, the physical world together with the virtual world becomes a mixed world. Items from the virtual world, such as the scroll, are mapped over the physical world.\\
	\scriptsize{
	\cite[Figure~2]{lankoski2004-songsnorth}, DOI: \href{http://dx.doi.org/10.1145/1028014.1028083}{10.1145/1028014.1028083}\\
	Copyright Association for Computing Machinery, Inc. Reprinted with permission.}
	}
	\label{figure:p413-lankoski-Figure-2}
\end{figure}

From \citeN{desouzaesilva2009-introduction}, it is obtained that pervasive games are persistent worlds (\ie persistent in the physical world~\cite{lankoski2004-songsnorth}), sharing the trait of a persistence with virtual worlds. 
This means that it could be beneficial ``to exploit the coextensive %
virtual world as a `behind the scenes' resource for coordinating and managing devices and interaction in the physical space''~\cite{greenhalgh2001} (see Figure~\ref{figure:p413-lankoski-Figure-2}). %
Because the feature set above includes a Virtual Game World with World Persistence, and a Shared Data Space with Data Persistence, a virtual world engine was chosen as being in the same `product line'~\cite{bass2013-inpractice} as a would-be pervasive games engine~\citePervasiveMOO. Many of the surveyed games in \PervasiveMOO featured a persistent virtual world, but not all \eg ARQuake. If virtual world persistence is not needed (only physical), a different engine might be sufficient to stage the pervasive game \eg ARQuake used a graphics engine~\cite{thomas2000-arquake}. It is assumed that the would-be pervasive games engine must support virtual persistence to be more general-purpose.
One looming problem remained; pervasive games do share the trait of a persistence with virtual worlds, but since there was no usable definition of a `virtual world'~\citeWorldDEF, how was it possible to be sure that the architecture for a would-be pervasive games engine does indeed overlap with that of a virtual world? The case study in Chapter~3 of \mbox{\PervasiveMOO} evaluated that indeed a virtual world engine can implement a pervasive game, but it could be beneficial to know to what extent other traits are shared.

\bigskip
\bigskip

\graphicspath{{includes/5_img/}} %

\chapter{Mixed Reality and Scalability}
\label{ch:combinedwork}

Using the definition for `virtual world' and the Triad Representational Framework from the previous Iteration III, in this chapter, pervasive games research is properly aligned with that of virtual worlds (rather than just on the persistence trait) and the mixed reality aspect of pervasive games is dealt with. Architectures for pervasive games are scaled up to distributed systems. And, considering the use of non-standard input devices in pervasive games and the rise of \IoT, Iteration IV explicates the problem of extending pervasive games into pervasive applications, incorporating IoT. %

\section{`Virtual World' Defined and Compared}
\label{section:vw-defined-overlapping}

\newcommand{\ZERO}{zero\xspace}
\newcommand{\ONE}{One\xspace}
\newcommand{\MANY}{one or more\xspace}

In \WorldDEF, a virtual world is defined as having the following properties: Shared Virtual Temporality~(\texttt{1T}, requiring Virtual Temporality), Real-time~(\texttt{Rt}), Shared Virtual Spatiality~(\texttt{1S}, requiring Virtual Spatiality and \ONE Shard), \ONE Shard\footnotemark\footnotetext{
``With shards, players are divided up into groups and assigned to a unique copy (a shard) of virtual space, with each shard handled by a different group of servers; players are prohibited from moving between shards''~\citeWorldDEF.
} (\texttt{1Sh}), \MANY Software Agents~(\texttt{A$_\PLUS$}, requiring Virtual Temporality), Virtual Interaction~(\texttt{I}, requiring Software Agents, \ONE Shard and \MANY shared abstractions of time), non-Pausable (\texttt{nZ}, requiring Real-time), Persistence~(\texttt{P}, world persistence (\texttt{wP}) and/exclusively-or data persistence (\texttt{dP})) and an \mbox{Avatar (\texttt{Av})}.
This led to the definition for a virtual world being:

\renewcommand{\ZERO}{\textsc{zero}\xspace}
\renewcommand{\ONE}{\textsc{one}\xspace}
\renewcommand{\MANY}{\textsc{many}\xspace}

\blockquote{\small{
	A simulated environment where:
	\MANY\footnote{one or more.} agents can virtually interact with each other, act and react to things, phenomena and the environment;
	agents can be \ZERO\footnote{exactly zero.} or \MANY human(s), each represented by \MANY entities called a \mbox{`virtual self'} (an avatar), 
	or \MANY software agents;
	all action/reaction/interaction must happen in a real-time shared spatiotemporal non-pausable virtual environment; 
	the environment may consist of many data spaces, but the collection of data spaces should constitute a shared data space, 
	\ONE\footnote{one and only one.} persistent shard.
}}

\let\ZERO\undefined
\renewcommand{\ONE}{One\xspace}
\let\MANY\undefined

\noindent Using this definition, it is possible to see if the software system architecture in each iteration can support a virtual world. This can give an indication of how one architecture has built upon knowledge of the previous.

\subsection{Is the GDD a Virtual World?}
\label{section:gdd-vs-virtual-world}

The GDD of \TheGDD, was designed as a web-based technology with similarities to a wiki. The GDD fails to support Virtual Temporality; any recorded times would be related to Simulated Temporality, so the GDD fails to support Shared Virtual Temporality. The GDD was designed to be readily available at all times for easy editing \ie not turn- or tick-based and therefore a Real-time technology. Because the GDD was designed as a document style tool, it fails to support Virtual Spatiality and, by consequence, Shared Virtual Spatiality. The node structure of the GDD was designed to be possibly highly distributed, but still \ONE Shard. No Software Agents were designed for and therefore Virtual Interaction was only between users \ie a communication technology. Since events were recorded according to Simulated Temporality, the non-Pausable criterium is easy to fill. World Persistence is also supported by the requirement of the GDD to be readily available at all times, albeit for designers of games behind a computer. Data persistence was an explicit requirement of the GDD, using revision control tracking and a backup system. Each user in the system would be given an entity to which multiple views and nodes could be linked, satisfying the Avatar property. The result for $\forall$\texttt{p}:\AND~is then negative: \texttt{\N \EQ VT[1T,A$_\PLUS$,I]\AND VS[1S]} (see Table~\ref{table:worlddeftest}) \ie failure to provide a shared virtual spatiotemporal environment for interaction with software agents. Rather than a virtual world, the GDD can be classified as equal to Google Docs~\mbox{\citeTheGDD}, a persistent communication technology.

\subsection{Is Traveur a Virtual World?}
\label{section:traveur-vs-virtual-world}

From Table~\ref{table:worlddeftest} it can be seen that Traveur scored similarly to the GDD with respect to being a virtual world; only satisfying one more property. Traveur also recorded times according to Simulated Temporality, failing to support Shared Virtual Temporality \ie since Traveur was a mash-up of technologies, it was easier to rely on real-world time than to synchronize a virtual time. Traveur was a Real-time technology, with the map functionality showing updates in the range of seconds. According the characteristics of \PervasiveMOO, it is not required that all pervasive games make use of Virtual Spatiality, but through the map functionality, Traveur supported Virtual Spatiality that was mapped to the physical world. Although the synchronization between the Streetside and Creator components was poorly implemented, they did form a shared data space \ie \ONE Shard. Software Agents were never implemented in Traveur; doing so would have been a reason to implement Virtual Temporality. Similar to the GDD, software agents were never implemented, so Traveur also fails to support Virtual Interaction. Simulated Temporality was used, rather than Virtual Temporality, so the non-Pausable property was trivial to fulfill. World Persistence was supported through the mobile phones providing ubiquity of access for the Parkour runners; although poorly designed, the architecture strived to maintained ubiquitous availability. Each of both Streetside and Creator had their own persistent database, satisfying the Data Persistence requirement. And, finally, each user in the system was assigned an entity in the Streetside mobile community, which was then used system wide \ie an Avatar. The result for $\forall$\texttt{p}:\AND~is negative: \mbox{\texttt{\N \EQ VT[1T,A$_\PLUS$,I]}} (see Table~\ref{table:worlddeftest}). In comparison to the GDD, Shared Virtual Temporality is implemented in the map functionality, but the lack of Virtual Interaction leaves Traveur insufficiently worldly. 

\begin{table}[t]
\centering
\ttfamily\selectfont
\begin{tabular}{|l|c|c|c|c|c|l|c|c|c|c|}

\textsc{technology}		& 1T
								& Rt
										& 1S    				
												& 1Sh   		
														& A$_\PLUS$ 
																& I		
																		& nZ
																				& P							
																						& Av
																								& $\forall$\texttt{p}:\AND	\\
\hline                                                                  
the GDD					& \R	& \Y	& \R 	& \Y	& \R 	& \R	& \Y	& \Y	& \Y	& \N						\\
Traveur					& \R	& \Y	& \Y 	& \Y	& \R 	& \R	& \Y	& \Y	& \Y	& \N						\\
pervasive engine		& \Y	& \Y	& \R 	& \Y	& \Y 	& \Y	& \Y	& \Y	& \Y	& \N						\\
\hline

\end{tabular}
\caption{Classification of the architectures according to \protect\WorldDEF} %
\label{table:worlddeftest}
\end{table}

\subsection{Pervasive Games Overlap with Virtual Worlds?}
\label{section:pervasivemoo-vs-virtual-world}

\WorldDEF determines a set of properties that a virtual world exhibits. If an engine features those properties, then a virtual world is supported by the engine \ie formal requirements can be formed on how to enable those features in the engine.
The choice of using a virtual world engine as a would-be pervasive games engine can be revisited by comparing the properties of a `virtual world' (in Section~\ref{section:vw-defined-overlapping}) with the desired characteristic feature set for a pervasive games engine (in Section~\ref{section:repurposing-vw-engine}) \ie it is possible to see what other properties they share besides persistence. 
Shared Temporality and Spatiality were explicitly named in the feature of a Virtual Game World with World Persistence; Virtual Temporality is implied, because Software Agents and persistence are also required (see further). However, a pervasive game world is said to overlap with both the virtual and physical, classifying a pervasive game as mixed reality, according to~\WorldDEF, \ie a pervasive game has Shared Spatiality, but not purely Shared Virtual Spatiality. The Real-time property is implicit in the pervasive game requirement for world persistence and ubiquitous availability \ie in a pervasive game the player can always access the game world, supporting temporal expansion. Again, the part of the game world player is accessing is not necessarily the virtual one; a distinction between pervasive games and pervasive computing ``lies in relation to ubiquity of access; pervasive computing without access to a computing device might prove difficult, but a player partially denied access to a computing devices can potentially still be in the game''~\citePervasiveMOO. The property of \ONE Shard is explicit in the characteristic of a Shared Data Space with Data Persistence', as is the property of Software Agents in the characteristic of Reconfiguration, Authoring and Scripting in Run-time. Interaction is mentioned in Virtual Game World with World Persistence, but a pervasive game is mixed reality \ie not all interaction is going to be virtual. The domain of \PervasiveMOO is Technology-Sustained Pervasive Games, so it can be understood that many actions/interactions in the physical world will be mapped to actions/interactions in the virtual world. The property of non-Pausable is not mentioned explicitly as a characteristic in the feature set, but is implied though world persistence and ubiquitous availability. Persistence is mentioned explicitly in the form of both world and data persistence. And, finally, if players of the pervasive game are to interact with the virtual world, they must have an entity ``through which all their in-world activity is channelled''~\citeWorldDEF \ie an Avatar. The result for $\forall$\texttt{p}:\AND~is then negative: \texttt{\N \EQ VS[1S]} (see Table~\ref{table:worlddeftest}) \ie a pervasive game is not guaranteed to support a virtual world due to the games's mixed spatiality.

An initial reaction to this negative result might be to conclude non-correlation (\ie pervasive games characteristics do not fully describe a virtual world), but this result clearly shows that the majority of properties defining a virtual world are characteristics that are preferable to a would-be pervasive games engine \ie a substantial improvement over choosing a virtual world engine as being in the same product line as a would-be pervasive games engine, based solely on the overlapping trait of persistence~\citePervasiveMOO. 
From the result it is possible to conclude that, because of the mixed nature of a pervasive game, not all pervasive games must support a virtual world, but they can. Considering the definition of `virtual world', the characteristic of a Virtual Game World with World Persistence of \PervasiveMOO should then be read as: a game world where some elements are virtual \ie not a virtual world \textit{per se}. As an example, Traveur was supposed to be a pervasive game;  Traveur could feature a virtual world, but isn't required to do so; if a virtual world is featured, the would-be pervasive games engine should be able to stage it. 

\subsection{One World or a World of Worlds?}
\label{section:delegation-vs-virtual-world}

In the face of scaling up architectures to accommodate a massive amount of interactions, \Delegation compares issues from the domain of {\MMOW}s with those from \IoT. It is a tautology that a \MMOW supports a virtual world, but it can be questioned whether architectures, for distributed pervasive applications incorporating \IoT, implement one virtual world or many. One important conclusion from \WorldDEF is that it clearly defines what constitutes a single virtual world or multiple, in the face of distributed (centralized and decentralized) systems. The property of \ONE Shard dictates that ``if a technology has more than one data space, those spaces must be merged at one point or another for them to be considered one virtual world; each other copy/shard is considered another world''~\citeWorldDEF. It is still to early to know if \IoT will be enabled by a single architecture or middleware, but there is a risk of `fragmentation'~\citeWorldDEF; it is this fragmentation that coincides with a negating of the \ONE Shard property, and in all likelihood the property of Shared Temporality as well.

\section{Mapping Between the Virtual and Physical}
\label{section:mapping-between-virtual-physical}

According to Section~\ref{section:pervasivemoo-vs-virtual-world}, the overlap between pervasive games and virtual worlds is clear; a pervasive games engine does not necessarily fully support a virtual world, because a pervasive game is mixed reality. It was particularly a lack of Shared Virtual Spatiality that caused a negative result \ie to what degree a pervasive game makes use of the physical or virtual world can vary, but a characteristic of a technology-sustained pervasive game is that the game world overlaps with both the virtual and the physical~\citePervasiveMOO. How then is a pervasive games engine supposed to handle this type of spatiality, where the physical can be mapped to the virtual and \viceversa? In Section~\ref{section:pervasivemoo-vs-virtual-world}, pervasive games are said to support Virtual Temporality, but temporality can also be mixed (see Section~5.3.5 in~\WorldDEF). How is this to be handled? In the domain of Pervasive Games, research exists tying the virtual to the physical, but an ``integrated model for dealing with space and time''~\citeTimeSpace is lacking. Such a model does exist in the domain of GIS, and since the domains of GIS and Pervasive Games overlap, the Triad Representational Framework can be exapted to pervasive games, as demonstrated in \TimeSpace. In combination with \cite{dix2005} and \cite{langran1992-timegis}, virtual time and space can then be mapped to the physical world and \viceversa.

Considering an overall conceptual representation for geographic phenomena, it can be assumed that any such representation is composed of entities, properties and relationships~\cite{peuquet1988-dual}. In a cartographic representation, two dominant views exist, the `geometric structure' (\eg size, shape, orientation, color, height and location) %
and the `graphic image' (\eg temperature or height of a particular location or relation to other locations). %
In geometric structure, the entity is a spatial object, whereas in the graphic image, the entity is a location.
Spatial objects are higher-order information, based on information from a location-based perception of the physical world. %
Although not entirely distinct, \citeN{peuquet1988-dual} presents these views in a unified Dual Model, on the account that ``neither view is intrinsically better than the other, but are logical duals of each other''; the geometric structure view is referred to as `object-based' and the graphic image view is referred to as `location-based'. 
If the object-based view is referred to as the \textsc{what}, and the location-based view as the \textsc{where}, then using the Dual Model, two categories of spatial queries can be formed: 
\begin{compactitem}
\item \textsc{what}~$\rightarrow$~\textsc{where} ~\eg given an object, where is it located?
\item \textsc{where}~$\rightarrow$~\textsc{what} ~\eg given a location, what objects are located there? 
\end{compactitem} 

\citeN{dix2005} have shown how virtual space can be mapped to and from physical space through `measured space', a representation of space captured in sensors (and actuators --
projection is given as an example of mapping directly from virtual space to the physical, but this author would argue that such a projection would also have to cross measured space \eg the projector also requires calibration). 
Combining Peuquet's %
Dual Model and \citeauthor{dix2005}'s three types of space, virtual objects (spatial object in object-based view) can be related to virtual locations (locations in location-based view) and mapped indirectly (through measured space) to physical locations, and \viceversa. If geometric structure is read by sensors, this can be mapped similarly to a virtual object.

Since pervasive games can be played in physical or real `world time'~\cite[p.34]{langran1992-timegis}, change must be accounted for \ie time must be added to the Dual Model. \citeN{peuquet1994-framework} extends her own Dual Model with a time-based view, forming the Triad Representational Framework. In the time-based view, the basic entity (in the overall conceptual representation) is a single unit of time. Using Triad, virtual time (time in the time-based view) can be related to virtual objects and virtual locations. 
If the time-based view is referred to as the \textsc{when}, the Triad framework enables spatiotemporal queries of the following forms:
\begin{compactitem}
\item \textsc{what}~\texttt{+}~\textsc{where}~$\rightarrow$~\textsc{when} ~\eg given an object at a particular location, when was the last time it changed or appeared?
\item \textsc{what}~\texttt{+}~\textsc{when}~$\rightarrow$~\textsc{where} ~\eg given an object in a time span, what trajectory through space did the object take?
\item \textsc{where}~\texttt{+}~\textsc{when}~$\rightarrow$~\textsc{what} ~\eg if at a particular location, what objects passed by after a particular time?
\end{compactitem} 
Virtual time can be considered an abstraction of time simulated by a computer system~\citeWorldDEF, and virtual time must be mapped to physical time. If virtual time is instrumented from physical time, similar to physical space above, then \cite{dix2005} can be applied to time \ie virtual time is indirectly mapped to physical time over `measured time'. \citeN[p.34]{langran1992-timegis} notes the difference between physical time and measured time, summarizing this concept in GIS literature.

Now that it has been shown how the Triad framework can be used to map the virtual to the physical and \viceversa, it must be clarified how the framework benefits a pervasive game. Literature on pervasive games from a technical perspective is scarce~\mbox{\citePervasiveMOO}; no literature was found by this author describing how to map the physical world, through sensors, to the internal representation found in pervasive game technology. GIS modeling is done as an information model, so the Triad framework has an object-based view that can be easily mapped to internal representation of game objects in the engine.
Each of the three views of Triad can be implemented using the overall conceptual representation of an entity with properties, connected to other entities via relation. Or, a more compact representation is possible, where an entity has location-based and time-based views stored in its properties.

Because Triad relates objects, locations and time, the framework also highlights redundancy or dependencies in an architecture~\citeTimeSpace. For example, in the Traveur architecture mentioned above: location data is generated through the map functionality, on the client side, and networked to the Streetside community server; users and their properties are object data in Streetside; skills (properties) for each user and the skills that can be performed at a particular location are found in Creator; and, time is found in relation to all data documenting when the data was recorded or changed. Querying for data is restricted to what is stored in each component. If Triad is not used, designers could be unaware that multiple copies of the same data are being stored in multiple components~\cite{peuquet1988-dual}, causing inconsistency. If Triad is used to relate the different views, dependencies can be drawn between the different components. Spatiotemporal queries such as: 
\begin{compactitem}
\item \textsc{what}~\texttt{+}~\textsc{where}~$\rightarrow$~\textsc{when}: \\
\indent 
``when was the last time a trainer with the jump skill was here?''
\end{compactitem}
will cause: all trainers (the object-based \textsc{what}) to be queried from Streetside; whether they have the jump skill  (properties pertaining to the objects) to be reconciled with Creator; and the locations (the \textsc{where}) of trainers, present and in the past (the \textsc{when}), to be queried from Streetside. It is also implicit, that consistency is maintained between Streetside and the client devices, with respect to the present and past locations of trainers. 
In \TimeSpace, Triad is applied to the pervasive game \CNH, demonstrating how the model is mapped to the internal representation of the game engine. Publications describing GIS logical and physical designs for pervasive games were left as future work.

\section{Scaling up the Architectures to Distributed Systems}
\label{section:scaling-architectures}

The requirements, design and development of the three software system architectures has been discussed. 
By aligning the domain of Pervasive Games with that of Virtual Worlds, it is possible to scale up to distributed architectures, using advances from the domain of Virtual Worlds.

\subsection{Client / Server}

A requirements analysis was performed for the new GDD medium and a model created, but no concrete system was architected. The most straightforward architecture to implement the GDD model would be that of client/server (see Figure~\ref{figure:GDD-architecture}); clients connecting to a single server through HTTP, with one or more redundant servers as backup (see Figure~\ref{figure:GDD-architecture}, redundant servers are not shown). Data persistence would be through the use of a database, and elementary interoperability through the specification of HTTP, as the communication protocol; HTML can play an important role in allowing the client to use dynamic content and interfaces~\citePervasiveMOO. The architecture of the GDD bares resemblance to that of \CNH (compare Figure~\ref{figure:GDD-architecture} with that of Figure~\ref{figure:CNH-architecture}), but without a sensory system on the client, for context-awareness, or a proxy component, for supporting for heterogeneous systems (outside of HTTP for interoperability). Since all transactions could be relayed through the centralized server, providing authority would be trivial. Scalability is problematic with a single server~\cite{yahyavi2013-p2pmmog}, handing perhaps hundreds or thousands of simultaneous users, but not more. Of course, cloud computing can be used in this scenario, but this is discussed in the next section.
The backup server must be kept up to date in the event of system failure, but consistency is trivial, since the active server always has the primary copy~\cite{yahyavi2013-p2pmmog}.

\figuremacro{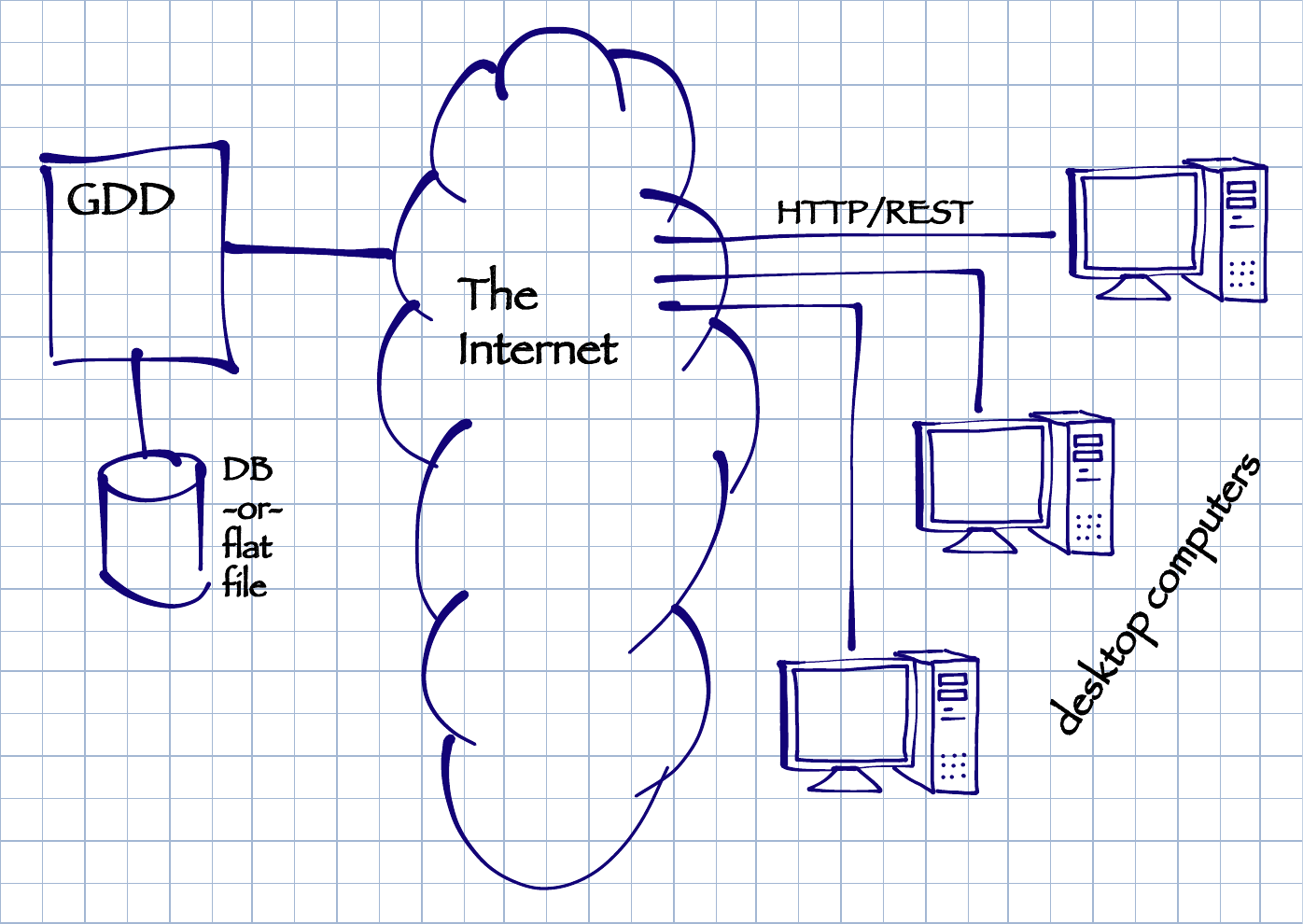}{GDD architecture}{One centralized GDD component (with a redundant backup) serving one or more desktop computers.}
\figuremacro{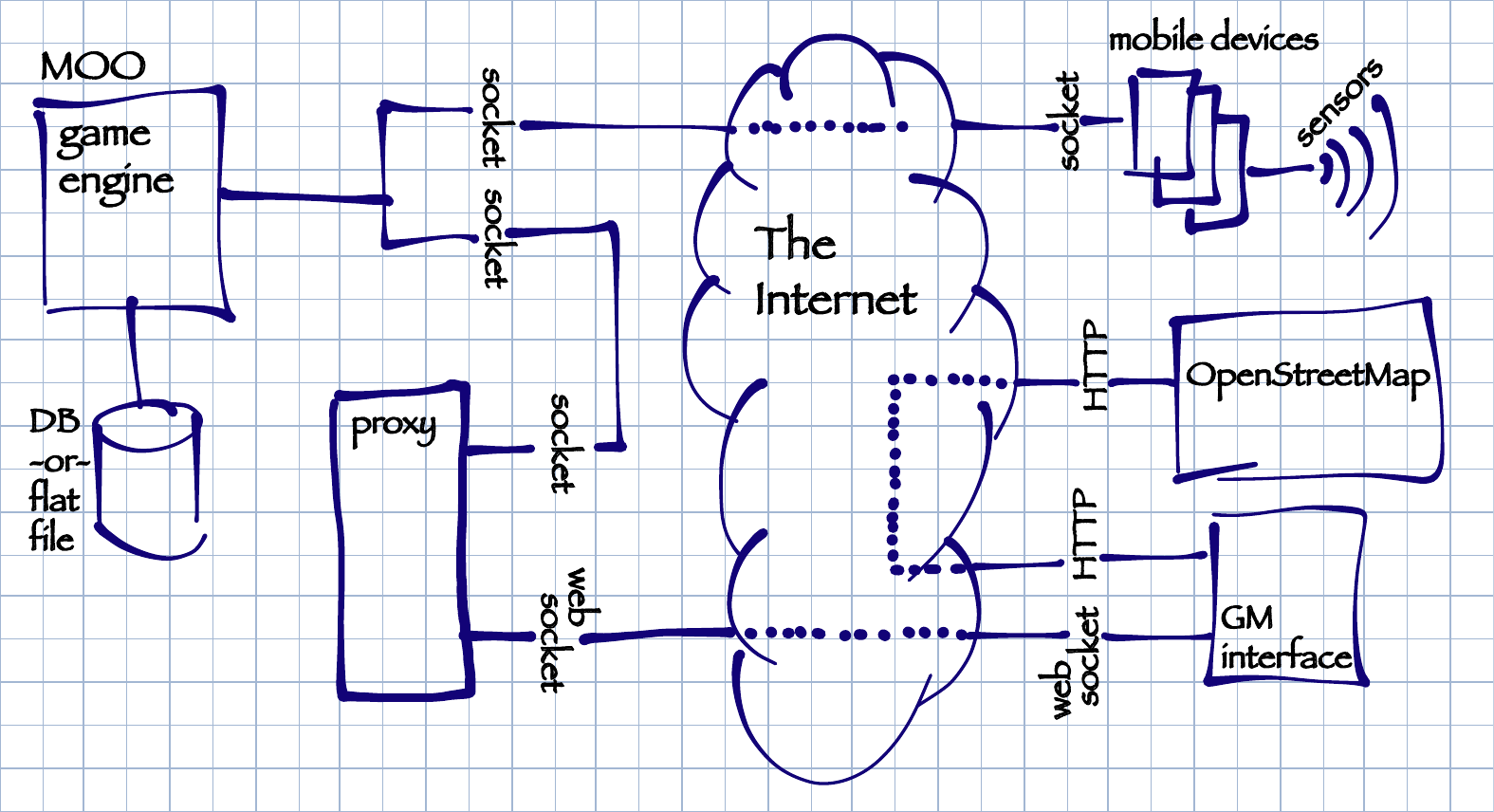}{\CNH architecture}{The MOO game engine component serving mobile devices with sensors, and a GM (game master) interface tool through a proxy object. This figure is a replication of Figure~3.2 of \PervasiveMOO; a detailed explanation of the architecture can be found in Section~3.3 of the book.}

If a sensory system were to be added to the GDD architecture, the architecture would start to resemble that of \CNH. The sensory system can be in the form of, for example, mobile phones with sensor hardware, wireless sensor networks, or service-oriented architectures providing external sensor data. If many types of sensors are used simultaneously, support for heterogenous devices and systems must be provided for; in the \CNH architecture, the proxy component (depicted in Figure~\ref{figure:CNH-architecture}) partly handled interoperability.

\subsection{A Centralized Server Cluster}
\label{section:centralized_server_cluster}

Although the MOO game engine, used in the \CNH architecture (see Figure~\ref{figure:CNH-architecture}), fulfilled many of the required characteristics as a would-be pervasive games engine, the component was outdated~\citePervasiveMOO. Since the domain of Pervasive Games has been aligned with that of Virtual Worlds, advances in virtual world architecture can be used to scale up pervasive games architectures to distributed engines. 
Virtual world engines are already in existence that use a centralized cluster of servers for load-balancing (\eg ~\citeN{bigworld2002})~\mbox{\citePervasiveMOO}, so the MOO game engine component in Figure~\ref{figure:CNH-architecture} can be replaced with a centralized cluster of servers to achieve cloud computing. This to an extent alleviates the scalability problem, but in exchange for communication overhead needed to coordinate the cluster~\cite{yahyavi2013-p2pmmog}.
To lower latency and bandwidth, improving scalability further, a hybrid solution can be considered \ie a peer-to-peer system in combination with a centralized server cluster (see Section~\ref{section:peer-to-peer}).

\subsection{Heterogeneous Clients}

Figure~\ref{figure:Traveur-architecture} depicts the final Traveur architecture, after the change of the dependency graph, from two direct dependencies to one direct and one indirect dependency (see Figure~\ref{figure:Traveur-dependency-graph} for the prior graph). 
Traveur highlighted that in a pervasive setting, ``heterogeneity is the norm'', stemming from the situation that the server components used in Traveur were heterogeneous. 
In actuality, the situation in Traveur could have still been worse. The Streetside community pages were designed to be available on a wide range of devices via HTTP and HTML, and the client was switched from Android OS to iOS, but only one single smartphone type was fully supported simultaneously \ie non-HTTP clients could also have been heterogeneous leading to more interoperability issues~\citePervasiveMOO. Heterogeneity of peers can be found in peer-to-peer networking (\eg peers can be chosen to be `superpeers', supporting higher responsibilities and with higher levels of resources~\cite{yahyavi2013-p2pmmog}), but in general high degrees of heterogeneity client side means more interoperability issues.

\figuremacro{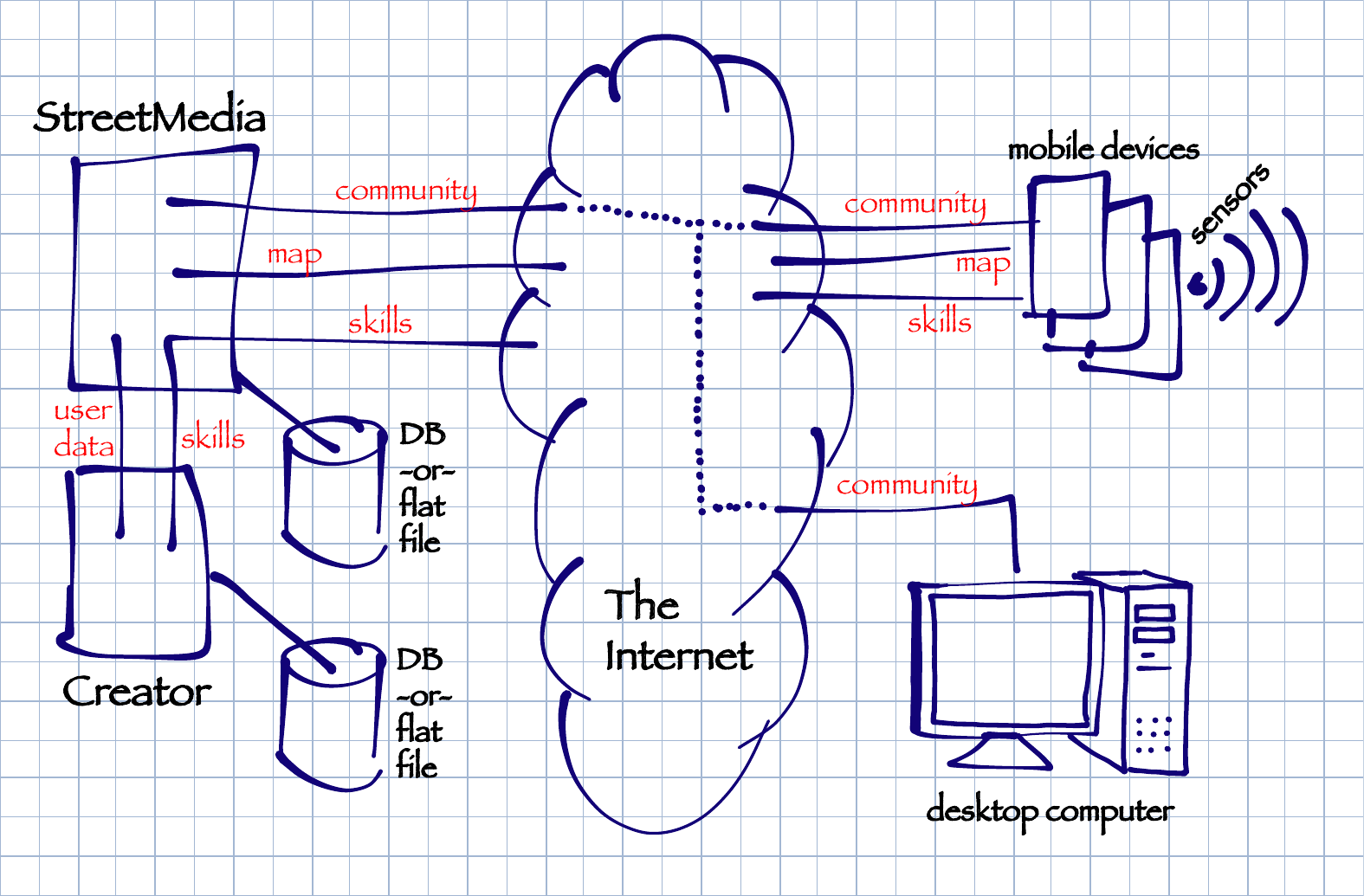}{Traveur architecture, after change of dependency graph}{The community function (which included user login data) was accessible through both mobile device and desktop computer, through HTTP and/or REST.}

\subsection{Heterogeneous Servers}
\label{section:heterogeneous-servers}

To make Streetside and Creator interoperable, each would have to behave as a service-oriented architecture. Streetside did provide REST~\cite{fielding2000-rest} interfaces and Creator supported modular adaptors to mitigate communication, but the services were not robust. The initial dependency graph, of each client having a direct connection to Streetside and Creator (see Figure~\ref{figure:Traveur-dependency-graph}, BEFORE), was changed on the account of login authentication, which was handled by Streetside. Creator could query Streetside for login authentication, but this could result in an odd situation where a user was logged in Creator, with the authentication server going down after login. Since Streetside and Creator were not in the same geolocation, querying from one server to the other also incurred latency. Changing the dependency graph (see Figure~\ref{figure:Traveur-dependency-graph}), with the client only needing a single connection to Streetside, resolved the odd authentication situation, but did not resolve the client's dependency on Creator, which was residing on another hardware server. At least with the updated dependency graph, the entire Traveur service could be deemed out of service, without incurring latency for authentication. 

Since both Streetside and Creator were designed to be stand-alone servers, replication services were not considered at design time. Consistency was implemented through a bulk transfer of changes (to user objects) to Creator, but since Streetside and Creator were not in the same geolocation, this transfer incurred large amounts of latency. A more efficient replication algorithm for user data would have to be implemented, but development resources in Traveur were already overused. An obvious solution would be to colocate both servers in the same geolocation, but this was denied by the stakeholders. 

Considering Traveur had heterogeneity server side, the architecture was more complex than that of \CNH; in the face of this complexity, a few different approaches could have been taken. First off, since Streetside and Creator both provided a unique part of the Traveur service, ideally both should have ensured availability. As mentioned in \PervasiveMOO, this requires ``reliability \ie mature systems, preventing failure by being fault tolerant and supporting recoverability''. Neither Streetside nor Creator were mature systems; each had been under development and not fully tested for stability (engine maturity was a primary reason for choosing a MUD-based engine for \CNH). When errors were encountered in Streetside or Creator, these caused system down time rather than less severe levels of failure. Also, in the case of system failure, no redundant backup servers were in place to maintain service. Since the Traveur client was directly or indirectly dependent on both components, the chances of having a degraded system were, of course, higher than if a single server were used.
As a solution, if only two physical hardware servers were available, two redundant copies of the same component could have been run on their respective servers, with a dynamic switch on failure. Or, as an alternative, a monitoring process could have been started to monitor each critical component for downtime, restarting the component in the event of failure. In the case of hardware failure, this still means a degraded system, since the redundancy and monitoring would be affected by the failure. A solution is to run a copy of each component on each server hardware, but this was denied due to the fact that each stakeholder wanted control over their own service. 

Although not encountered at the time, Traveur suffers the same scalability problem as \CNH \ie a single server supporting a limited amount of users. The same solutions apply here as discussed above, in Section~\ref{section:centralized_server_cluster}.
If both clients and servers are heterogenous, then essentially the system could be consider to be that of a peer-to-peer system or a hybrid architecture~\cite{yahyavi2013-p2pmmog}.

\subsection{Peer-to-Peer}
\label{section:peer-to-peer}

As mentioned in Section~\ref{section:centralized_server_cluster}, scalability can be improved by moving towards a peer-to-peer system in combination with a centralized server cluster (see Figure~\ref{figure:Hybrid-architecture}) \ie allowing clients to communicate directly with each other rather than having to communicate indirectly over the server. But, care must be taken to consider security and cheating, so that clients don't interrupt information dissemination, perform illegal actions, or gain access to unauthorized information~\cite{yahyavi2013-p2pmmog}. 
It is possible to move away from a centralized server cluster entirely, distributing the authority of the game, but then security and cheating becomes even more of an issue. 
The survey of Chapter~2 in \PervasiveMOO includes a decentralized architecture for pervasive games, but security and cheating are noted as open research questions in Chapter~4 of the same book.

\figuremacro{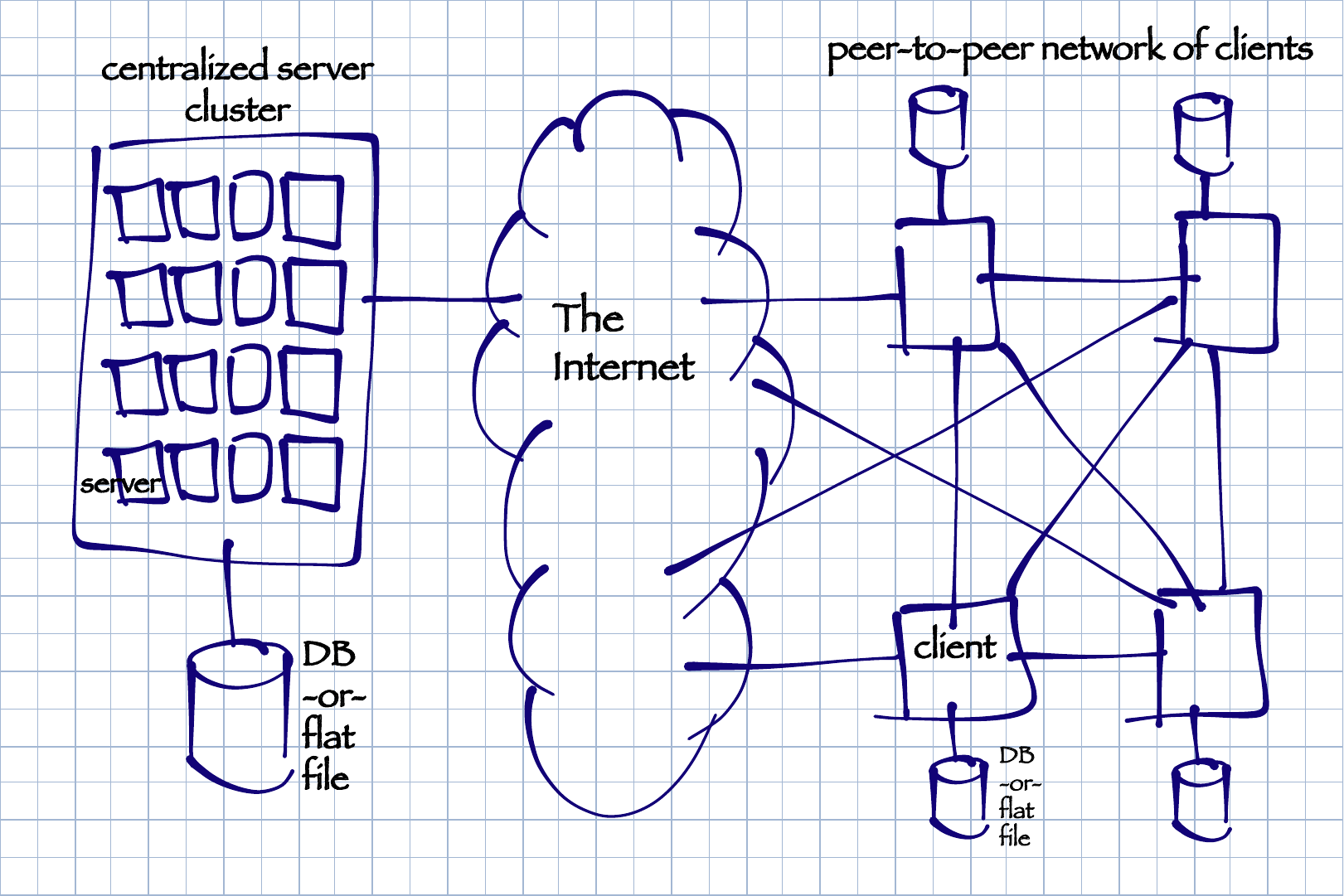}{Hybrid architecture}{A centralized server cluster, with servers (depicted as squares) inside the cluster, serving a peer-to-peer network of clients.}

Another large consideration is that in a mobile setting, devices have limitations in connectivity, latency, memory, and processing power. \citeN{yahyavi2013-p2pmmog} state that this makes it ``unlikely that mobile platforms will become powerful peers in a massively multiplayer game'' (\eg a massively multiplayer online [virtual] world), but do add that peer-to-peer mechanisms might be advantageous in a local setting. This brings the discussion up to par with resource availability, in an \IoT setting (see Section~\ref{section:availability}).

%

%

%
%

\section{Distributed Pervasive Applications, Incorporating \IoT}
\label{section:delegation}

Pervasive games need a sensory system for context-awareness (which is usually made up of heterogeneous devices~\citePervasiveMOO) and part of \IoT could serve as such a sensory system \ie pervasive games will also have to contend with \IoT. Considering game engines can be repurposed~\citePervasiveMOO, it is possible to talk about a broader set of pervasive applications, enabled by pervasive games engines with support for \IoT.

With the rise of the Internet of Things, research is ongoing to design and develop a platform that can enable \IoT~\citeDelegation. Research on {\MMOW}s (\ie large scale virtual worlds) includes various distributed architectures to deal with the issue of scalability, %
having already encountered issues that are now also surfacing in the domain of \IoT. \Delegation surveys the domain of {\MMOW}s for properties that are affected by scaling an architecture, and then evaluates how these are dealt with in the domain of \IoT, noting the similarities and dissimilarities. In has been shown in previous sections that the domain of Pervasive Games overlaps with the domain of Virtual Worlds, allowing for advances in the domain of Virtual Worlds to be reapplied \eg how to deal with properties of scale.

In \Delegation, six properties related to scaling have been discussed, with one conclusion being that ``{\MMOW}s can clearly learn from advances, in resource utilization, availability, and responsiveness with respect to (partially) disconnected networks, from the domain of \IoT''. Pervasive games are scaled up using research from the domain of {\MMOW}s, so the shortcomings in these areas are discussed in the following sections.

\subsection{Resource Utilization}
\label{section:resource_utilization}

One of the main tenets of \IoT is the support for sensors and mobile devices having limited computing power \ie `smart objects'~\cite{lopez2011-resource,miorandi2012-adhoc}. Support for smart objects is not common in the domain of {\MMOW}s, so research on such objects can't be directly adopted by pervasive games from the domain of Virtual Worlds, but the domain of Pervasive Games does have its own research with respect to Heterogeneous Devices and Systems (see Section~\ref{section:repurposing-vw-engine}) \eg a mechanism is needed so that the virtual can be mapped to the physical and \viceversa. Service-oriented architectures is an approach being used in the domain of \IoT to support interoperability between various platforms and systems~\cite{atzori2010-iot,miorandi2012-adhoc}, which is also part of Heterogeneous Devices and Systems for pervasive games. Interoperability is an open issue in both \IoT~\cite{gubbi2013-vision,miorandi2012-adhoc,atzori2010-iot} and Pervasive Games~\citePervasiveMOO. 
\citeN{lopez2011-resource} argue that \IoT shall ``aim to provide an architecture that will make use of the strengths of both, SOA [service-oriented architectures] and agent-based systems''; if pervasive games adopt an architecture from the domain of Virtual Worlds, then agents are already supported (see Section~\ref{section:vw-defined-overlapping}).

\subsection{Availability}
\label{section:availability}

The domain of {\MMOW}s has a considerable amount of research towards peer-to-peer and hybrid solutions~\citeDelegation, as discussed in Section~\ref{section:peer-to-peer}, but rarely contends with (partially) disconnected networks. 
Because \IoT must content with (partially) disconnected networks~\cite{lopez2011-resource}, pervasive games more closely relate to \IoT than {\MMOW}s in this respect. In the event of a disconnected state, remote resources (\eg cloud computing) are often unavailable, in contrast to local resources which might still be available, depending on the architecture \eg if the authority over local resources is a remote centralized one, then authorization over local resources cannot be obtained in a disconnected state and access to local resources is denied. For {\MMOW}s, a centralized architecture is often preferred giving the game developer more control over the architecture~\cite{yahyavi2013-p2pmmog}, but if pervasive games are to support (partially) disconnected networks or peer-to-peer, this will require the ``rethinking of game platforms and game engines''\footnote{as Theo Kanter always says.}.

\Delegation states that ``for consistency and security, there seems to be advances in both domains that can cross over to the other domain''; it is precisely in a decentralized scenario where there is much activity in the domain of \IoT that can possibly be adopted by {\MMOW}s, and by extension pervasive games. 

\subsection{Responsiveness}
\label{section:responsiveness}

``To achieve the appropriate responsiveness, \IoT is said to require two classes of traffic: the throughput and delay tolerant elastic traffic; and, the bandwidth and delay sensitive inelastic (real-time) traffic''~\citeDelegation. Considering real-time data ``is one of the properties that distinguishes {\MMOW}s from other distributed systems''~\citeDelegation, %
pervasive games relate more closely to \IoT with respect to responsiveness \eg in \PervasiveMOO, the use of `delay-tolerant' networking is mentioned as a solution in partially disconnected networks.

%
%

%

%
%

%
%
%

%

%
%

%

\graphicspath{{5_evaluation/figures/}} %

\chapter{Evaluation} %
\label{ch:evaluation}

In the previous two chapters, it has been shown how work on the GDD and Traveur influenced the design and development of a would-be pervasive games engine. Research on pervasive games has been aligned with that of virtual worlds, and it has been demonstrated how to map the virtual to the physical, and \viceversa. Pervasive games have been scaled up to distributed architectures using research on {\MMOW}s. And, finally, research from \IoT was used to explicate the problem for future research on distributed pervasive applications. 

In this chapter, the research question and methodology of the dissertation is first evaluated, followed by an evaluation of the work from Chapter~\ref{ch:architectures} and \ref{ch:combinedwork}.

\section{Validating the Research Question}

The entire research question follows from the primary question of: \textit{can engine technology from the domain of Computer Video Games be repurposed to stage technology-sustained pervasive games?} That engine technology from domain of Computer Video Games can be repurposed has already been established in literature~\cite{lewis2002}. Before expanding to pervasive applications, the domain of this dissertation is limited to Technology-Sustained Pervasive Games; computer video games are also technology-sustained and so share this trait with pervasive games. The similarity of the two domains is exemplified by the statement of \citeN{montola2009} that technology-sustained pervasive games are ``computer games interfacing with the physical world''.

If the shared activity of gaming is looked at, a similar overlap can be observed. In the Pervasive Discourse by \citeN{nieuwdorp2007-discourse}, she explains that the first time the word `pervasive' was used in conjunction with `gaming' was in relation to LARP (live action role-play), of which MUD was considered a `virtual counterpart'. MUD is both a type of game as well as a virtual world.

\section{Verifying the Research Methodology}
\label{section:verify_research_methodology}

In order to qualify as Design Science research, instead of Systems Development, projects must fulfill three requirements stated verbosely in the beginning of Chapter~\ref{ch:methodology}. They can be summarized as:
\begin{inparaenum}[(i)]
\item projects make use of rigorous research methods;
\item the knowledge produced has to be related to an already existing knowledge base; and, 
\item new results should be communicated to both practitioners and researchers.
\end{inparaenum}
An outline of the research methodology and strategies has been presented in Chapter~\ref{ch:methodology}. The knowledge produced in this dissertation is related to the existing knowledge base presented in Chapter~\ref{ch:relatedwork} and to the related work presented in each individual publication included in this dissertation. Results have been communicated to the research community through publication and to practitioners through research collaboration \eg Street Media of the Traveur project (see Section~\ref{section:traveur}). %

\section{Evaluating the Design of the Architectures}
\label{section:evaluting_design_architectures}

Evaluation of each publication is done within each publication itself, but the work, presented in Chapter~\ref{ch:architectures}, of how each system is said to build on previous iterations can also be evaluated.

\subsection{The GDD as a Basis for Pervasive Applications}

To illustrate how the GDD formed a basis for later work, the requirements of the GDD (in Section~\ref{section:architecture-thegdd}) can be compared with the desired component features for pervasive games (in Section~\ref{section:architecture-pervasivemoo}). 
With respect to the target audience for the GDD (which were individuals working with desktop class computers), the virtual environment was required to be ``readily available at all times to all users'', which is exactly the requirement for World Persistence in \CNH~\citePervasiveMOO. In order to support ``collaborative user editing'', users must have access to the same Shared Data Space. The specified `revision control' and `backup system' for the GDD provides Current and Historical [Game] State of the system and also the necessary Data Persistence. Again with respect to the target audience, a web-based approach was suggested so as to be able to support all users on Heterogeneous Devices and Systems. The GDD was required to feature Roles and Permissions for editors. And, finally, a Bidirectional Communication was imperative for the GDD, for both user communication and update dissemination.

An important contribution of the GDD, which is still not widely adopted as of this writing, is the `view' concept; this concept returns in \CNH, where a custom interface can be considered for each participant role in a pervasive game (see Section~2.3.5 of \PervasiveMOO), and leads to an open issue (see Section~4.1.4 of \PervasiveMOO). The GDD proved to be a persistent communication technology (see Section~\ref{section:gdd-vs-virtual-world}), the properties of which influenced the forming of a definition for a `virtual world' in \WorldDEF.

\subsection{The Heterogeneity of a Pervasive Service}

The GDD was not pervasive (only supporting temporal expansion for the target audience), so Traveur was the first pervasive project that was done under this author's PhD studies. Before shifting to a pervasive service, Traveur was a game. To evaluate that Traveur was a basis for the pervasive games architecture of CNH, it is possible to examine exactly which properties or features were inherited by \CNH.

Out of the characteristic feature set of \PervasiveMOO (see Section~\ref{section:repurposing-vw-engine}), Traveur to a certain degree featured all of them. A Virtual Game World was provided for, but World Persistence was poor due to the instability of the system. Data Persistence was not an issue, there was no loss of data, but the algorithm ensuring consistency of the shared data space was slow, incurring latency in the system. Traveur as a whole was only available on a single platform, the iOS mobile platform, but the HTTP interface provided by Streetside was with the intent of supporting a wide variety of devices \ie there was little support for Heterogeneous Devices. Traveur was ambitious with regard to Context-Awareness, making use of a wide variety of location dependent and biometric data. Support for Roles, Hierarchies and Game Master Intervention is evident in the training system employed by Traveur; trainers with special privileges would be the control mechanism, to make sure that trainees had completed the proper training, before moving on to the next level. Current and Historical Game State was available in both Traveur as a whole and in the map functionality. Because Traveur was based on Creator, some Runtime Authoring and Scripting was provided for in the game related part of the project. Since Streetside was still under development, Streetside did not offer any support for scripting at the time. Since Traveur ran on a mobile phone, the phone itself could serve for Bidirectional Communication. The Traveur client served as a unidirectional Diegetic Communication channel, in the sense that the community function, provided by Streetside, showed the activity of other participants in real-time. The ability to leave comments in the map function was also appropriated as a sort of slow Bidirectional Diegetic Communication channel, as well. 

The design of the map functionality for Traveur (see Section~\ref{section:map_functionality}) fed into the client, server and WebMap tool, for \CNH \eg dealing with position localization, real-time updates, uncertainty, and the visualization of game state on a WebMap. The concept of ``heterogeneity being the norm'' influenced \PervasiveMOO, stimulating research into the challenges surrounding interoperability.
That two stakeholders in the project disputed over control of their segment of the data, fed into \Delegation the notion of authority.

\subsection{Evaluating the Feature Set for Pervasive Games}

In \PervasiveMOO, a component feature set for a would-be pervasive engine was derived; each feature is verbosely summarized  in Section~\ref{section:repurposing-vw-engine}. Initially, pervasive games are shown to overlap with a virtual world based only on the shared persistence trait \ie a virtual world engine was chosen to be in the same product line as a would-be pervasive games engine. In Chapter~3 of \PervasiveMOO, the component feature set and the choice of a virtual engine as pervasive engine are verified through the case study of the pervasive game called \CNH. 

\section{Evaluating the Combining of Results}
\label{section:evaluting_combined_results}

In Chapter~\ref{ch:combinedwork}, publications are used in combination with one another to handle mixed reality and scalability; this must also must also be evaluated. 

\subsection{Overlap of Pervasive Games and Virtual Worlds}
\label{section:overlap-pervasive-virtual}

Since no usable definition existed for a `virtual world', properties of a virtual world were obtained through grounded theory in \WorldDEF, and used to form a definition. In Section~\ref{section:pervasivemoo-vs-virtual-world} above, the obtained definition is used to determine if pervasive games support a virtual world. The overlap is such that all properties of a virtual world are shared by a pervasive game, except for the mixed reality aspect of pervasive games. 
To verify that the properties of a virtual world (summarized in Section~\ref{section:vw-defined-overlapping} above) are actually supported, the \CNH architecture (used in the case study of \PervasiveMOO) can be examined to see if all properties are indeed supported. 

\paragraph{Shared Virtual Temporality (\texttt{1T})} %

The \CNH architecture makes use of the LambdaMOO (MOO) game engine, a MUD-based engine (see Chapter~3 in \mbox{\PervasiveMOO}). According to \mbox{\WorldDEF}, in order to support Shared Temporality, an architecture must support Virtual Temporality. MOO offers access to Simulated Temporality, but it is the programmer's responsibility to create `timers'~\cite{hess2003-yibco} for software agents that require Virtual Temporality. Considering \CNH used a centralized architecture, creating Shared Temporality was trivial; timers were created centrally in the game engine and the output disseminated to all clients. \CNH was temporality expanded~\citePervasiveMOO and Simulated Temporality was used to record when certain events had entered the system \eg last known player position or when a skill had been performed. No attempt was made to double check the accuracy of the Simulated Temporality; it was assumed clocks on individual devices were synchronized using Coordinated Universal Time. Since \CNH made use of both Simulated and Virtual Temporality, \CNH is temporal mixed reality. 

\paragraph{Real-time (\texttt{Rt})} 

All devices were able to communicate with the MOO engine in real-time. 
The MOO architecture was built to handle multiple concurrent connections, via the Telnet protocol, in a Virtual Terminal program~\mbox{\citePervasiveMOO}. In \CNH, Telnet in combination with MUD Client Protocol was utilized for all connecting devices; clients connected directly to MOO and were responsible for keeping the connection alive for real-time communication. To enable the WebMap game master tool, a proxy object~\mbox{\citePervasiveMOO} was used to keep the connection alive, since the tool itself opened one or more connections on demand instead of a constant connection.

\paragraph{Shared Virtual Spatiality (\texttt{1S})} %

Because pervasive games are spatially expanded, making use of mixed reality, pervasive games (\ie \CNH) are \textbf{not} required to support Virtual Spatiality (see Section~\ref{section:pervasivemoo-vs-virtual-world}). The notion of space in MUD comes from narrative and a room-based topology~\citeWorldDEF. In \CNH, the room-based topology was repurposed for the questing system, and a GPS-based coordinate system used as a spatial representation instead. The players were not given any notion of space beyond the physical \ie the client directed them through physical space, but without the use of a visual map functionality or similar. It was only the WebMap tool that provided a real-time visual map for game masters.

\paragraph{One Shard (\texttt{1Sh})}

Similar to the Shared Temporality property, this property is trivial due to the centralized architecture of MOO; all clients can cache data, but the authority and primary copy holder of all objects is the single server. Clients can not function without a valid connection to the server.

\paragraph{Software Agents (\texttt{A$_\PLUS$})} %

To support Software Agents, an architecture needs to support Virtual Temporality~\citeWorldDEF; which \CNH does indeed, see above. All objects in the MOO engine have the potential to be Software Agents through scripting. A primary example of this is the \texttt{void\_walker} NPC, which mimicked a real \CNH player~\citePervasiveMOO.

\paragraph{Virtual Interaction (\texttt{I})} 

To support Virtual Interaction, an architecture needs to support both One Shard and one or more shared abstractions of time~\citeWorldDEF. \CNH satisfies the One Shard property (see above) and supports both Shared Simulated and Virtual Temporalities \ie abstractions of time.
In order for a technology to support Virtual Interaction, users must be able to interact with both other users and the world~\citeWorldDEF. As a diegetic communication channels, in \CNH, players were able to interact with each other both through message passing and proximity. But, ``using the client as an non-diegetic bi-directional communication channel was decided against, as not to break the mythos of the game''~\citePervasiveMOO. Due to limited development resources, no non-diegetic communication was created, so this defaulted to email and phone conversations \ie interaction, but not virtual. To interact with the world, players could act and react to physical objects which had a virtual counterpart, through the game client and QR-Codes. Or interact with the \texttt{void\_walker} mentioned above, which mimicked a player~\citePervasiveMOO.

\paragraph{non-Pausable (\texttt{nZ})} %

To be non-Pausable, the architecture needs to support the Real-time property~\citeWorldDEF, which \CNH indeed does, see above. \CNH was temporally expanded, so the game world was available to any player at any time \ie the virtual game world could not be paused risking not being available.

\paragraph{Persistence (\texttt{P})} %

The Persistence property requires an architecture to support both world persistence and data persistence. The underlying characteristic for persistence is based on the work by Richard Bartle, maker of the original MUD1~\citeWorldDEF, so stating that MUD supports persistence would be a tautology. And, the architecture for \CNH, being MUD-based, means it also supported both world and data persistence.

\paragraph{Avatar (\texttt{Av})}

Each user of \CNH (including game masters) was assigned a player entity, as per %
the MOO engine, through which their virtual interactions where channeled. Because the MOO engine asserted that each user was to only have a single identity, the MOO engine did not support crossmedia. To allow game masters to be logged in to the system via two different devices, simultaneously, a `ghost' player was created to maintain the second log on~\mbox{\citePervasiveMOO}. The uniqueness constraint for an avatar has been removed in \WorldDEF; without it crossmedia can be supported.

\paragraph{\phantom{....}}

By evaluating the \CNH architecture against the properties of a virtual world, the result is the same as shown for a pervasive game in Table~\ref{table:worlddeftest}. This leaves the mixed reality nature of a pervasive game to be evaluated.

\subsection{Evaluating the Virtual / Physical Mapping}

In Section~\ref{section:mapping-between-virtual-physical} it has been shown that using \cite{dix2005} and \cite{langran1992-timegis}, in combination with \cite{peuquet1994-framework}, physical time and space can be mapped to virtual time and space, and \viceversa. Since both pervasive games and GIS make use of the Earth's geography, \TimeSpace presents an exaptation of Triad to pervasive games.
To evaluate the mapping of physical to virtual, the architecture of \CNH can be examined for the relations present in the Triad framework~\cite{peuquet1994-framework} and three types of space \cite{dix2005}. The first part of evaluating this mapping is done in \TimeSpace; the article evaluates  that Triad is applicable to pervasive games using \CNH. This leaves the usage of \cite{dix2005} and \cite{langran1992-timegis}, in combination with Triad, to be evaluated. 

In \CNH, the object-based view of the \textsc{what} consists of game objects (organized in the taxonomic MOO hierarchy), as well as various groups of objects administered by the specialized \texttt{generic\_admin\_group}. The time-based view of the \textsc{when} consists of snapshots of events at regular intervals. And, the location-based representation of the \textsc{where} consists of GPS measurements~\citeTimeSpace. Objects in the MOO hierarchy and groups (the \textsc{what}) were either created during the development of the game, or in run-time \eg the basic \texttt{game\_master} profile was created during development, but the virtual counterparts, of crowd-sourced player generated artifacts, were created in run-time. When a player starts their game client for the first time, they are asked to create a profile. By creating a profile, a virtual player object is created and assigned a place in the MOO hierarchy. %
After profile creation, the player's location can already be read from the player's device;
sensors on the player's device pick up GPS signals and measure longitude, latitude and possibly time~\cite{nco2013-gps}.
The physical location (the \textsc{where}) of the player is measured via sensors, constituting measured space \ie the mapping of physical space to measured space. %
Readings from sensors can be saved to device storage, with possible truncation or modification at this stage or each subsequent stage of storage \ie saved readings in the virtual space are mapped to measured space.
The time of a GPS reading (the \textsc{when}) can be obtained either directly from the GPS signal, or in association with the current time of the device, which is commonly synchronized with Coordinated Universal Time \ie simulated computer time mapped to physical time via measurement. In \CNH, there are many simulated times (at least one for each device), which must be compared with the authority of time, namely the server. Virtual time is the abstraction of simulated time in the server. 
Once location and time are on device, they are transmitted via mobile network to the game engine, where they are associated with the object of the player that is logged in \ie all three views of the \textsc{what, when} and \textsc{where} are related to each other in the game engine. The game engine holds the primary copy of the entities, with `replicas'~\cite{yahyavi2013-p2pmmog} maintained by the game clients. 

During execution of the game, margins of error can form in stored measurements \eg due to limitations of the internal representation, transmission, rounding or truncation error. (It is also possible that a game master had to pick a location on the map, but the GPS location entered was only marginally accurate.) These inexact measurements will be mapped back to physical space, offset from the intended location. The WebMap game master interface prototype of \CNH visualized player positions in the game world; using current and historical data \ie GPS traces could be drawn on the interface map. The WebMap connects simultaneously with both  the game engine and OpenStreetMap~\cite{osmf2004-map}; the game engine for the game data and the OpenStreetMap for map data. Replicas from all three views of the \textsc{what, when} and \textsc{where} are transmitted via networking to the system hosting the WebMap. Entities from the views are moved from virtual space into measured space, the map display. For a single player, a line of dots can be placed on screen, with each dot marked with the player's ID and a timestamp \ie object-based view (\textsc{what}) and time-based view (\textsc{when}). The location of the dot on the map represents location-based view (\textsc{where}). Mapping between measured space and the physical is done when the game master looks for the players represented on the map in physical space, and finds that they are, or are not, where intended, depending on the margins of error in the system. 

This completes the evaluation of Triad in combination with \cite{dix2005} and \cite{langran1992-timegis}, having explained how physical time and space is measured through sensors, moved into measured space, into the virtual, and \viceversa. 
In combination with the previous Section~\ref{section:overlap-pervasive-virtual}, the overlap of pervasive games with virtual worlds and also the mixed reality nature of pervasive games has been evaluated.

\subsection{Scaling Up Architectures and Incorporating \IoT}

It is advantageous to align the domain of Pervasive Games with the domain of Virtual Worlds, so that advances in the aligned domain can be adopted. Otherwise, pervasive games would be left reinventing existing technologies. 
In Section~\ref{section:scaling-architectures}, it was shown how the outdated architecture for \CNH could be scaled using advances in architecture from the domain of Virtual Worlds. These advances are grounded in existing research.

In the beginning of Section~\ref{section:delegation} it is stated that part of \IoT could potentially be used as the sensory system for a pervasive application. To evaluate this, a prototype dubbed IoMOOT was built using the \CNH engine, and subsequently demonstrated at VIP Day 2012, hosted by this author's research group at the time. 
A small scale model of a residential home was used as the physical space for the demonstration. The internal spatial representation of MUD was used to simulate the space of the residential home virtually. Inside the model home, various `things' were present (\eg a lamp, fan and thermometer), each of which was given a virtual entity in the \CNH engine. The user of the system was also given a virtual entity in the engine, moving around virtual space relative to the users movement in physical space; the position of the user was simulated rather than using indoor position localization. The game client of \CNH on iOS was replaced by a TouchOSC~\cite{hexler2009-touchosc} interface, and the WebMap, which served as game master interface, was replaced by a web page displaying the status of the home \eg lights on/off, fan on/off and current temperature.
Via mobile device a user could use the TouchOSC interface to control the devices in the home remotely and via the web to check the status of the home. The status web page was also available via any web enabled device or system. 
The device abstraction layer provided by MUD Client Protocol~\citePervasiveMOO, made it possible to easily replace the two user interfaces and provide connectivity to the \IoT enabled home. IoMOOT allowed the user ``to exploit the coextensive %
virtual world as a `behind the scenes' resource for coordinating and managing devices and interaction in the physical space''~\cite{greenhalgh2001}; %
this time applied to \IoT rather than pervasive games.

\Delegation surveys the domain of {\MMOW}s (\ie large scale virtual worlds) for properties that are affected by scaling an architecture, and then evaluates how these are dealt with in the domain of \IoT. Since the domain of Pervasive Games is aligned with that of Virtual Worlds, \Delegation is used to explicate the problem of pervasive games having to contend with scale and incorporate \IoT (see Section~\ref{section:delegation}). Considering game engines can be repurposed~\mbox{\citePervasiveMOO}, it is possible to talk about a broader set of pervasive applications. Explicating the problem, for distributed technology-sustained pervasive applications, is the start of Iteration IV of the method framework for Design Science Research. The rest of this iteration is left as future work (see Chapter~\ref{ch:futurework}).

\graphicspath{{5_conclusion/figures/}} %

\chapter{Conclusion} %
\label{ch:conclusion}

Currently, there are no reusable game engines available for pervasive games; without such engines, developers would be left continually reinventing existing technologies. If technology-sustained pervasive games can be understood as computer games interfacing with the physical world, the first part of the research question of this dissertation asks \textit{can engine technology from the domain of Computer Video Games be repurposed to stage technology-sustained pervasive games?}

Several related works suggest frameworks and middleware for pervasive games (i.e., custom-built solutions), but the solutions are not compared with existing game engine technology (more `general-purpose' solutions \cite{gregory2009}, and only a few publications repurpose an existing game engine in their project. \citeN{paelke2008-lbgaming} state that many pervasive games ``implement their own custom game engine by tying together available standard components for distributed applications and filling the gaps with custom code''. Although a viable solution, ``tying together available standard components'' does not provide any foresight into how the architecture will behave with respect to scalability. ARQuake was developed by repurposing the original Quake engine, as mentioned by \citeN{lewis2002}. Because a persistent virtual world was not required by the game, ARQuake could be implemented by repurposing a ``graphics engine rather than a virtual world engine. But, without a persistent virtual world, ARQuake's times of play (when the player is in the game world) can not be temporally expanded. In the Ambient Wood~\cite{thompson2003} a MUD-based virtual world engine (with support for persistence) was chosen to stage the game. There is no mention in the literature of possibly reusing the architecture of Ambient Wood to stage additional pervasive games, or if the architecture could be scaled to support more players. \citeN{kasapakis2015-trends} state that ``the game engine organization model is largely dictated by the game scenario to be supported'' and this author agrees. If a general-purpose pervasive games engine is to be created, commonalities between game technologies must be found. 

This dissertation contains the design and development of three software system architectures and clearly showing how each influenced the requirements for technology-sustained pervasive games (see Chapter~\ref{ch:architectures}). 
The GDD, in \TheGDD, proved to model a persistent communication technology, of which the requirements analysis, design and `view' concept (see Section~\ref{section:architecture-thegdd}) influenced Traveur and \CNH. Properties of a persistent communication technology influenced the definition of a `virtual world' in \WorldDEF (see Section~\ref{section:gdd-vs-virtual-world}). 
Traveur highlighted that ``heterogeneity is the norm'' and served as exploratory research into pervasive and context-aware computing. A detailed account of the Traveur architecture, and its connection to \mbox{\Acrobats}, has been presented in Section~\ref{section:architecture-traveur}, the requirements analysis and design of which influenced \CNH. \PervasiveMOO contains an extensive systematic review into pervasive games, resulting in a feature set describing characteristic features of a would-be pervasive games engine. Using the feature set, a virtual world engine was chosen as being in the same product line as a would-be pervasive games engine, based on the shared trait of a persistence (see Section~\ref{section:repurposing-vw-engine}).

A contribution of \PervasiveMOO is that it highlighted that a model, mapping time and space from the physical to the virtual, and \viceversa, was missing and that no usable definition for `virtual world' existed. \WorldDEF provides a definition for a virtual world and it is used to show that domain of Pervasive Games overlaps with that of Virtual Worlds, except for a discrepancy based on mixed reality (see Section~\ref{section:vw-defined-overlapping}); %
that the majority of properties defining a virtual world are characteristics that are preferable to a would-be pervasive games engine. This leaves only the mixed reality nature of a pervasive game to be dealt with. To handle the mixed reality nature of pervasive games, the Triad Representational Framework~\citeTimeSpace is combined with \cite{dix2005} and \cite{langran1992-timegis} to map time and space, from the virtual to the physical, and \viceversa (see Section~\ref{section:mapping-between-virtual-physical}). 

Since computer game engines \textbf{can} be used to stage a pervasive game, the research question continues by questioning whether \textit{advances from the domain of Computer Video Games can be used to scale up pervasive games to distributed systems}. It has been shown that domain of Pervasive Games \textbf{does} overlap with that of Virtual Worlds, and by aligning the domains advances in architecture, from the domain of Virtual Worlds, are used to scale up pervasive games architectures to distributed systems (see Section~\ref{section:scaling-architectures}). In a recent publication by \citeN{kamarainen2014-cloudgaming}, the subject of cloud computing is discussed as a possible solution for pervasive and mobile computing. They remark that latency is the ``main challenge'' for cloud gaming, with most interactive games requiring response times that only ``local deployment scenarios'' can deliver. As a solution, they ``propose to use [a] hybrid and decentralized cloud computing infrastructure, which enables deploying game servers on hosts with close proximity to the mobile clients''.

The second part of the research question asks: \textit{considering the use of non-standard input devices in pervasive games and the rise of Internet of Things, how will this affect the architectures supporting the broader set of pervasive applications?} To exploit local resources, the Fun in Numbers (\FiiN) platform (presented in \PervasiveMOO) features a distributed multi-tiered (\ie four layered) large-scale architecture~\cite{chatzigiannakis2011-implement}. The bottom layers of the \FiiN architecture enable support for ad-hoc networks and \IoT. In the domain of \IoT, ``endeavors already underway in an attempt to create an \IoT platform, but a solution that addresses all the aspects required by the \IoT is yet to be designed''~\citeDelegation. The contribution of \Delegation is that it surveys the domain of {\MMOW}s for properties that are affected by scaling an architecture, and then evaluates how these are dealt with in the domain of IoT. %
Using results from \Delegation pervasive games are extended to pervasive applications, incorporating \IoT; properties pertaining to scalability from the domain of {\MMOW}s and IoT are compared and the result used to explicate the problem of scaling architectures for pervasive applications (see Section~\ref{section:delegation}). \Delegation contains open issues for {\MMOW}s and IoT, which extend to pervasive games. In an early publication, \citeN{broll2009-perci} describe the Perci Framework, which connects mobile devices (via a proxy) to a web server, forming a pervasive service that leverages \IoT. The Perci Framework can be used by games or other applications, and is in the domain of Ubiquitous Computing. The open issues and the possible overlap between the domain of Pervasive Games and Ubiquitous Computing leads to the future work presented in Chapter~\ref{ch:futurework}.  

Evaluation of the dissertation is done in Chapter~\ref{ch:evaluation}. Evaluating the design and development of the three software system architectures is done in Section~\ref{section:evaluting_design_architectures}; Chapter~3 of \mbox{\PervasiveMOO} is the evaluation of
the characteristic feature set and the choice of a virtual engine as pervasive engine, verified through the case study of \CNH. 
The combined results of this dissertation are evaluated in Section~\ref{section:evaluting_combined_results}; \TimeSpace evaluates that the Triad Representational Framework is applicable to pervasive games. Advances, from the domain of Virtual Worlds, used to scale up pervasive games to distributed systems are grounded in existing research.

The aim of this dissertation has been to allow advances from the domain of Computer Video Games to be used to scale up pervasive games to distributed systems. And, considering game engines can be repurposed, incorporating \IoT, explicate the problem for distributed technology-sustained pervasive applications. The implication of this dissertation is to ensure that pervasive games are not left reinventing existing technologies.

\graphicspath{{5_futurework/figures/}} %

\chapter{Future Work} %
\label{ch:futurework}

The most obvious future work, is the design and development of an architecture for distributed pervasive applications. Chapter~4 of \PervasiveMOO outlines open issues for a would-be pervasive games engine. And, \mbox{\Delegation} and \mbox{Section~\ref{section:delegation}} explicit the problem of scaling such an architecture to incorporate \IoT. 
The following points then summarize the other direct extensions of this dissertation:

\begin{itemize}
\renewcommand{\labelitemi}{$\square$}

\item In \PervasiveMOO, distributed and decentralized architectures is already listed as a challenge for pervasive games engines. Such architectures have the potential to improve scalability, but security and cheating is far more problematic than for a centralized architecture. Ad-hoc networks are also touched upon in \PervasiveMOO, but the concept is generalized to (partially) disconnected networks in \Delegation. If resources are to be available and accessible in a disconnected state, then this will require the rethinking of game platforms and game engines \eg with respect to authority, consensus and consistency.  

\item The extension of ubiquitous computing is mentioned in \PervasiveMOO and sometimes as a vision for \IoT~\cite{gubbi2013-vision}. Internet of Things has the potential to offer more context-awareness through sensor technology, while actuators can enable technology to control more of the physical environment. The complexity of dealing with a massive amount of context-information in combination with the existing entities and agents in the game engine must be dealt with; this opens up avenues to research in the areas of Ambient Intelligence~\cite{callaghan2014-intellmmo,xiao2015-lic} and Big Data~\cite{zaslavsky2012-bigdata}.

\item \citeN{rehm2015-mediator} state that the ``rather intangible idea of the IoT, generating greater value and service by enabling data exchange between manufacturers, service providers and other connected devices (`things'), has recently been replaced by the broader, more tangible, concept of Cyber-Physical Systems''. Similar to how a virtual world is used as a `behind the scenes' resource for pervasive applications in the dissertation, \citeN{rehm2015-mediator} ``believe that virtual worlds can serve as platforms to facilitate the integration required by Cyber-Physical Systems''. This new domain boosts feedback systems that integrate computation, networking, and physical processes, through embedded systems \ie real-time computing.

\item If decentralized architectures are to succeed, mechanisms must be found to provide for security and privacy \eg authority and data integrity. In \PervasiveMOO, anti-cheating is mentioned, but in \IoT this can be generalized to just security and privacy.

\item Interoperability is already mentioned as an issue in \PervasiveMOO, but the use of \IoT makes the issue even more important. Pervasive games are usually under the control of a limited number of stakeholders, so interfacing between heterogeneous systems is often manageable. \mbox{\Delegation} mentions the possibility that \IoT will fragment into many different platforms. 
If that is the case, then accessing \IoT will require the interfacing with many platforms; this includes having to deal with the various classes of responsiveness mentioned in \Delegation (see Section~\ref{section:responsiveness}).

\item Research in the domain of Pervasive Games has been aligned with that of Virtual Worlds. Pervasive game architectures have been scaled up to distributed systems, and the problem of scalability has been explicated for distributed pervasive applications, incorporating \IoT (see Section~\ref{section:delegation}). It might be interesting to survey the domain of Pervasive and Ubiquitous Computing to compare the architectures found there to existing game engines.

\item If \IoT does fragment into many different platforms: a fragmented platform structure plus technology-sustained pervasive games, aligned with virtual worlds, resembles the current Metaverse~\cite{dionisio2013-metaverse} concept (which includes pervasive concepts such as augmented reality). There are various definitions of the Metaverse~\cite{frey2008-solipsis,rehm2015-mediator}, but a fragmented platform structure resembles the system of interconnected virtual worlds, described by~\citeN{frey2008-solipsis} \ie an internet of virtual worlds.

\end{itemize}

\backmatterSU

\begin{scriptsize} %

\renewcommand*{\bibfont}{\small}

\defbibfilter{technologies}{
  type=software or
  type=technology
}
\printbibliography[title=\textsc{Technologies},filter=technologies]
\printbibliography[title=\textsc{References},nottype=technology]

\end{scriptsize}
\cleardoublepage
\phantom{.}
\cleardoublepage


\section*{Supplementary material}

**DEPARTMENT OF COMPUTER AND SYSTEMS SCIENCES**
Stockholm University/KTH
*www.dsv.su.se*
**Ph.D. theses:**

No 91-004 **Olsson, Jan**
An Architecture for Diagnostic Reasoning Based on Causal Models
No 93-008 **Orci, Terttu**
Temporal Reasoning and Data Bases
No 93-009 **Eriksson, Lars-Henrik**
Finitary Partial Definitions and General Logic
No 93-010 **Johannesson, Paul**
Schema Integration Schema Translation, and Interoperability in Federated Information Systems
No 93-018 **Wangler, Benkt**
Contributions to Functional Requirements Modelling
No 93-019 **Boman, Magnus**
A Logical Specification for Federated Information Systems
No 93-024 **Rayner, Manny**
Abductive Equivalential Translation and its Application to Natural-Language Database Interfacing
No 93-025 **Idestam-Almquist, Peter**
Generalization of Clauses
No 93-026 **Aronsson, Martin**
GCLA: The Design, Use, and Implementation of a Program Development
No 93-029 **Boström, Henrik**
Explanation-Based Transformation of Logic programs
No 94-001 **Samuelsson, Christer**
Fast Natural Language Parsing Using Explanation-Based Learning
No 94-003 **Ekenberg, Love**
Decision Support in Numerically Imprecise Domains
No 94-004 **Kowalski, Stewart**
IT Insecurity: A Multi-disciplinary Inquiry
No 94-007 **Asker, Lars**
Partial Explanations as a Basis for Learning
No 94-009 **Kjellin, Harald**
A Method for Acquiring and Refining Knowledge in Weak Theory Domains
No 94-011 **Britts, Stefan**
Object Database Design
No 94-014 **Kilander, Fredrik**
Incremental Conceptual Clustering in an On-Line Application
No 95-019 **Song, Wei**
Schema Integration: - Principles, Methods and Applications
No 95-050 **Johansson, Anna-Lena**
Logic Program Synthesis Using Schema Instantiation in an Interactive Environment
No 95-054 **Stensmo, Magnus**
Adaptive Automated Diagnosis
No 96-004 **Wærn, Annika**
Recognising Human Plans: Issues for Plan Recognition in Human - Computer Interaction
No 96-006 **Orsvärn, Klas**
Knowledge Modelling with Libraries of Task Decomposition Methods
No 96-008 **Dalianis, Hercules**
Concise Natural Language Generation from Formal Specifications
No 96-009 **Holm, Peter**
On the Design and Usage of Information Technology and the Structuring of Communication and Work
No 96-018 **Höök, Kristina**
A Glass Box Approach to Adaptive Hypermedia
No 96-021 **Yngström, Louise**
A Systemic-Holistic Approach to Academic Programmes in IT Security



\end{document}